\documentclass[aps,pra,showpacs,twoside,twocolumn,10pt]{revtex4-1}
\usepackage[colorlinks=true, citecolor=red, urlcolor=blue ]{hyperref}
\usepackage{times,epsfig,amssymb,amsfonts,amsmath,bm,subfigure,mathtools,amsthm,braket,soul,enumitem,color,physics,graphics,graphicx}
\UseRawInputEncoding
\usepackage[normalem]{ulem}
\usepackage{comment}
\usepackage{epstopdf}
\usepackage{multirow}
\usepackage{cancel}
\usepackage{soul}
\usepackage{xcolor}

\newcommand{\sh}[1]{{\color{black}#1}}
\newcommand{\sm}[1]{{\color{black}#1}}
\newcommand{\sud}[1]{{\color{black}#1}}

\begin{document}

\title{Imperfect  entangling power of quantum gates}
\author{Sudipta Mondal$^{1}$, Samir Kumar Hazra$^{2}$, Aditi Sen (De)$^{1}$}
\affiliation{$^1$ Harish-Chandra Research Institute,  A CI of Homi Bhabha National Institute, Chhatnag Road, Jhunsi, Prayagraj - $211019$, India\\
\(^2\) MURTI - Quantum Information Lab and Department of Mathematics, Gandhi Institute of Technology and Management, Bengaluru-562163, Karnataka, India }

\begin{abstract}

Achieving perfect control over the parameters defining a quantum gate is, in general, a very challenging task, and at the same time, environmental interactions can introduce disturbances to the initial states as well. Here we address the problem of how the imperfections in unitaries and noise present in the input states affect the entanglement-generating power of a given quantum gate -- we refer to it as imperfect  (noisy) entangling power. We observe that, when the parameters of a given unitary are chosen randomly from a Gaussian distribution centered around the desired mean, the quenched average entangling power -- averaged across multiple random samplings -- exhibits intriguing behavior like it may increase or show nonmonotonic behavior with the increase of disorder strength for certain classes of diagonal unitary operators. For arbitrary unitary operators, the quenched average power tends to stabilize, showing almost constant behavior with variation in the parameters instead of oscillating. Our observations also reveal that, in the presence of a local noise model, the input states that maximize the entangling power of a given unitary operator differ considerably from the noiseless scenario. Additionally, we report that the rankings among unitary operators according to their entangling power in the noiseless case change depending on the noise model and noise strength.

\end{abstract}

\maketitle

\section{Introduction}
\label{sec:intro}

Quantum tasks might potentially outperform classical protocols when certain quantum resources present either in the physical system or in the operations or in both are exploited.
The most prominent of these are quantum entanglement \cite{Horodecki2009}, quantum discord \cite{Modi2012, Bera_2018}, quantum coherence  \cite{coherenceRMP} which are  recognised as essential ingredients in numerous quantum information protocols, ranging from the transfer of both classical and quantum information between locations via the shared entangled state \cite{bennett1992, bennett1993} to the implementation of quantum gates associated with the building of a quantum computer \cite{nielsen_chuang_2010}. Furthermore, several experimental setups demonstrate how entangled pairs can be generated, such as through spontaneous parametric down-conversion in photons \cite{Zukowski12}, manipulating the internal energy levels of individual ions \cite{HAFFNER2008155}, and controlling the microwave pulses that interact with the superconducting qubits \cite{superconductingrev04}.

Quantum gates represented by unitary operators and the input states on which the gates operate are the necessary components for constructing an operational quantum circuit. To execute a protocol effectively in a circuit, quantum gates must be combined in such a way that the resource like entanglement necessary for that protocol is created. Therefore,  assessing the entangling capability of quantum gates is fundamental in quantum information processing and quantum computation \cite{Zanardi00, KrausCirac2001, Zanardi01, leifer01, wang02, Whaley03, Sudbery05, Bala09, Bala10, Linowski2020}. In order to measure this, the entangling power of a unitary operator is defined as the maximum entanglement of the output state produced following the action of the operator on the product input states (see Ref. \cite{Galve13} for other resource generating power of quantum gates).   Instead of maximal entangling power, the average entangling power of a quantum gate can be defined by averaging across a manifold of product states distributed according to some probability distribution \cite{Zanardi01}. \sud{Note here that instead of entangling power, one can also consider the gate fidelity, which is a crucial metric for assessing the performance of quantum gates \cite{Nielsen02, 10.5555/1972505,Sanders2015}. However, there are distinct differences between these two quantities -- the gate fidelity is evaluated after implementing the gate to assess its performance, while the entangling power of a quantum gate is determined during the circuit design, particularly when a task demands a certain amount of entanglement as a resource, and hence the focus lies on the properties of the output state rather than the state itself. In this work, we concentrate on the latter in the presence of imperfection and noise influencing the circuit.} In particular, when the system is in contact with the environment, it typically loses their coherence or become entangled with their surrounding environment, leading to the loss of quantum properties  and the breakdown of delicate quantum superpositions \cite{noise1, noise3, noisebook}. As a result, errors enter the system, reducing the computational capability of quantum algorithms or compromising information transfer, thereby diminishing the capacity of quantum communication protocols
and poses a significant challenge to the preservation and utilization of quantum resources, including entanglement.   In a similar vein, quantum operations employed to manipulate the quantum states are not always flawless. Therefore, it is critical to analyze the impacts of decoherence (imperfections) on the systems in order to develop strategies such as error correction \cite{nielsen_chuang_2010}, error mitigation \cite{McClean23errormitigation},  decoherence-free subspaces \cite{decoherence-free},   dynamical decoupling \cite{Lidardyndecoup}, reservoir engineering \cite{reservoirengineer96}, and identification of channels  \cite{discofreeze1, discofreeze2, discofreeze3, entfreeze, titfreeze}  to limit and overcome the obstacles. 

\begin{figure}
\includegraphics [width=\linewidth]{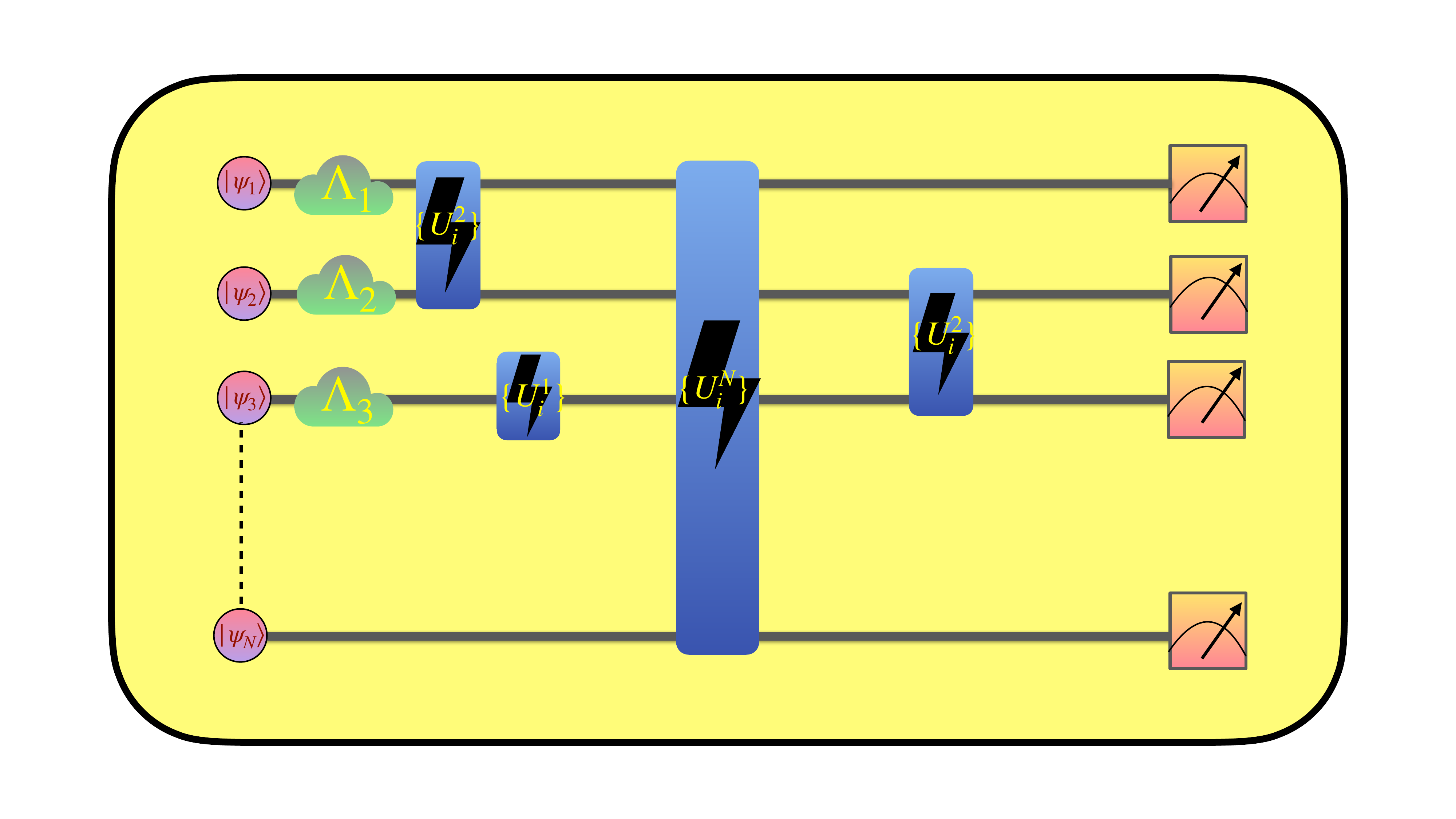}
\caption{{\bf Schematic diagram of the imperfect circuit.} Suppose \(|\psi_i\rangle\)s are the inputs prepared for different unitary operators  \(U^k\) with the superscript \(k\) (\(k=1,2\ldots N\)) being the number of qubits on which it acts. The initial states are chosen in such a way that after the action of  \(U^k\)  on it, the resulting state contains maximum entanglement that is possible to produce via \(U^k\). We consider two defects. (1) We assume that after the preparation of inputs and before the action of unitary, the input state gets disturbed by local noise, \(\Lambda_i\). We demonstrate that for a given unitary operator, the input state required to produce maximum entanglement in the presence of noise is different than the one in the noiseless scenario. Therefore, our results indicate that if the apriori noise model active in the circuit is known, the initial states have to be changed to generate maximum resources in the circuit. (2) We also consider the scenario where a given unitary operator is imperfect. In other words, the parameters, \(\eta\), \(\eta'\), and so on, in \(U^k (\eta, \eta', \ldots, )\)  cannot be tuned accurately due to faulty device. We choose these parameters randomly from some probability distribution. In this case, entangling power is computed by performing averaging over all such random configurations.}
\label{fig:schematic}
\end{figure}

We address the decoherence issue within the context of quantum circuits by raising the question -- ``How does decoherence affect the entangling power of quantum gates and consequently, quantum circuits?'' Defects can be introduced into this scenario in two ways: either the specified unitary transformation is inaccurate, or the input states are influenced by noise.    We investigate how these two forms of shortcomings have an impact on the entangling power of a given quantum gate, which we refer to as \sm{``imperfect (noisy) entangling power"} (see Fig. \ref{fig:schematic}).    It should be noted that the entangling power of operators has been investigated in the literature when the set of optimization is substituted by a set of mixed separable states rather than a set of pure product states, \cite{entpowermixedGuo14} or when the operation is noisy nonunitary  \cite{Liang15nonunit}. However, the noisy entangling power presented here differs from the previous studies in two ways. Firstly, the unitary transformation in our case remains unitary;  the parameters involved in the unitary operator only deviate from the intended one, i.e., if the parameters involved in the unitary operator are not precisely calibrated, they are selected from the Gaussian distribution with the mean being the desired value and \sm{a suitable} standard deviation (SD). The corresponding entangling power is estimated by choosing parameters randomly with the same mean and SD multiple times and then averaging them -- this is known as quenched average entangling power \cite{PhysRevLett.62.2503,disorderbook, disorderBerarakshit14}. Secondly, the maximization in entangling power is still performed over the set of pure product states which are subsequently affected by some local noisy channels. \sud{Moreover, note that our analysis is restricted to multiqubit systems, although it has been shown that the set of operations, thereby the performance of a certain task like circuit synthesis for transmons \cite{PRXQuantum.4.030327,PhysRevX.13.021028}, can be enhanced by considering qudits, especially in the presence of leakage, crosstalk, and errors.} 

We identify specific classes of diagonal unitary operators with nonvanishing entangling \sm{power} in which, when parameters are not precisely tuned, the quenched average entangling power can increase as well as exhibit nonmonotonic behavior (increase as well as decrease) as the disorder strength increases. Against the natural \sm{expectation} that defects impair a device's functionality, our results reveal the constructive impacts of disorder which have also been observed in diverse contexts in the literature \cite{bazaes2021effect, Pritam_2022, anuradha2023production}. In this work, we prove that the optimal input product state for a class of single-parameter diagonal unitary operators remains the same for the entire range of the parameters, implying that the input state providing maximal entangling power of a fixed unitary operator does not change due to imperfections. When parameters involved in arbitrary two-qubit unitary operators are chosen randomly, the quenched average entangling power becomes almost constant, and independent of the parameters while it oscillates with the variation of the parameters in the ideal situation.

On the other hand, if the circuit is influenced by noise, the resource that the gate can generate reduces, thereby affecting the performance. We illustrate that when the noise that is active \sm{in} the circuit is known, it is advantageous to adjust the input state so that the entangling power of the unitary operator is maximized. Specifically, we observe that the local depolarizing channels can have a major impact on the entangling power of an arbitrary quantum gate in four and higher-dimensional Hilbert space while the amplitude damping has \sm{the} least. 
Interestingly, we find that the strength of noise can modify the hierarchy among the unitary operators according to their entangling power and the multipartite optimal input state maximizing the power, demonstrating varying degrees of robustness in operator entanglement against noise.

The paper is structured as follows. In Sec. \ref{sec:frame}, we show how to adapt the well-known notion of entangling power, when noise acts on inputs or a given unitary for which entangling power has to be computed is imperfect. Sec. \ref{sec:imperfectunitarytwoq} studies the entangling power when imperfect unitaries act on two-qubit inputs while in Sec. \ref{sec:noisyentpower2qu}, the similar studies are carried out when the inputs are affected by local noisy channel although the unitaries are perfect. In Sec. \ref{sec:imperfection3qubits}, we go to a multipartite scenario where more than two qubits are chosen as inputs which can be influenced by local noise and the corresponding perfect as well as imperfect unitaries act on them. We provide concluding remarks in Sec. \ref{sec:conclu}.

\section{Entangling power framework in presence of imperfections}
\label{sec:frame}

Here, we introduce the setup for the noisy entangling power of quantum gates.  Let us first define the entangling power of a quantum gate, \(U\), in a noise-less scenario. In a two-party system,  the entangling power of  a given unitary operator, \(U\), is obtained by maximizing the entanglement produced after its action  upon the set of  product states \cite{Zanardi00, Zanardi01, wang02}, i.e., 
\begin{eqnarray}
    \mathcal{E} (U) = \max_{S}   [\mathcal{E}(U(|\psi_1\rangle \otimes |\psi_2\rangle))],
\label{eq:entpower}
\end{eqnarray}
where $S$ is the set containing product states  \(|\psi_1\rangle \otimes |\psi_2\rangle\), and \(\mathcal{E}\) is a valid entanglement measure \cite{Horodecki2009}. Estimates of such entangling power already exist in the literature,  however, they are considered only under the ideal scenario where the system is without any imperfection or decoherence. \sud{Note that instead of maximization, an alternate way can also be taken to define entangling power of \(U\) as \( e_p(U) := \overline{\mathcal{E}(U |\psi_1\rangle \otimes |\psi_2\rangle)}^{\psi_1, \psi_2}\), where the overline denotes the average over all \textit{product states}, \( |\psi_1\rangle \otimes |\psi_2\rangle \), distributed according to some probability density \( p(\psi_1, \psi_2) \) over the manifold of product states. Moreover, as mentioned in the introduction, one can also consider the fidelity of the gate after implementing the gate to determine its performance instead of entangling power. Mathematically,  gate fidelity is defined as the overlap between the desired output state and the state  obtained after applying the given gate \( \hat{U}_{\text{actual}} \)  compared to the ideal gate \( \hat{U}_{\text{ideal}} \) \cite{nielsen_chuang_2010}, i.e.,
\( F_{\text{avg}}(U_{\text{actual}}, U_{\text{ideal}})
    = \int d\psi \, \bra{\psi} U_{\text{ideal}}^\dagger U_{\text{actual}} \ket{\psi} \bra{\psi} U_{\text{actual}}^\dagger U_{\text{ideal}} \ket{\psi}\).
}

In this work, we focus on the maximum entangling power given in Eq. (\ref{eq:entpower}). Also, in practice, during the implementation of quantum gates, imperfections are inevitable, and hence it is crucial to incorporate the effects of decoherence in the assessment of entanglement-generation capability of quantum gates. There are two major ways via which decoherence can disturb the performance of the circuit (see Fig. \ref{fig:schematic}), they are as follows: \\
(1) when the unitary operator in action is imperfectly tuned, and \\
(2) when noise acts on the input product states, especially on the optimal input state which maximizes the entangling power of a given unitary operator.  \\

In both situations,  since the intended resource required for a certain task is not generated, the execution of the circuit does not lead to success. Let us elaborate on how to redefine the entangling power of a quantum gate by taking the above points into account.

{\it (1) Imperfection in unitaries.}
Let us now determine the response of the entangling power of quantum gates against imperfection present in the unitary.
Suppose, the goal is to implement \(2^k\)-dimensional unitary operator, 
$U^{k}(\eta_1,\eta_2, \ldots, )$ with \(\eta_1, \eta_2, \ldots \) being the parameters of the unitary which are fixed to a  value for a certain task. In an ideal case, the parameters can be exactly set to the values that are required while in the imperfect case,   \(\eta_i\)s may be chosen with some errors. To model these imperfections, let us choose \(\eta_i\)s involved in \(U^k\) randomly from some probability distribution with mean, \(\langle\eta_i\rangle\)  having some deviation. In particular, we choose the desired parameters \(\{\eta_i\}\)s (\(i=1, 2, \ldots\)) from the Gaussian distribution, \(G(\langle \eta_i\rangle, \sigma_{\langle \eta_i \rangle})\) with mean 
\(\langle\eta_i\rangle\) and the standard deviation (SD) \(\sigma_{\langle\eta_i\rangle}\) where the SD represents the strength  of the imperfection. The question is how to estimate the entangling power of quantum gates in this non-ideal scenario. 
Towards handling this situation, one can compute the entangling power of a quantum gate described in Eq. (\ref{eq:entpower}) for each random unitary configuration,  and then perform averaging over all such possible realizations. Such averaging is known in the literature as quenched averaging \cite{disorderbook, disorderBerarakshit14}. Mathematically, the quenched average entangling power can be represented as
\begin{eqnarray}
    &&\mathcal{E}^{avg}(U^k(\langle\eta_{1}\rangle, \sigma_{\langle\eta_1\rangle},\langle\eta_{2}\rangle, \sigma_{\langle\eta_2\rangle}, \ldots)=\int\mathcal{P} (\eta_{1}) \mathcal{P}(\eta_{2}) \ldots \nonumber \\
    && \times \mathcal{E}(U (\eta_{1},\eta_{2},\ldots)) d(\eta_{1}) d(\eta_{2})\ldots,
    \label{eq:avg_ggm}
\end{eqnarray}
where $\eta_{1},\eta_{2}, \ldots $ are the set of parameters which are chosen from the probability distribution $\mathcal{P}(\eta_{1})$, $\mathcal{P}(\eta_{2}) \ldots$ with mean \(\langle \eta_i\rangle\) and \(\sigma_{\langle\eta_i\rangle}\) (\(i=1,2, \ldots\)). Notice that when the faulty unitary acts on the product input state, the output state in each realization remains pure, thereby making the evaluation of entangling power easy while averaging. 

{\it (2) Noise in state-preparation. }
Let us first consider the two-qubit inputs on which four-dimensional unitary operators, \(U(4)\) act. In the case of inputs being affected by noise, the entangling power can be redefined as  
\begin{eqnarray}
\mathcal{E}^{\Lambda} (U) = \max_{S} \mathcal{E} [U(\Lambda^{(p)}(|\psi_1\rangle \langle \psi_1| \otimes |\psi_2\rangle \langle \psi_2|))U^{\dagger}],
\label{eq:noisyentpower}
\end{eqnarray}
where \(\Lambda^{(p)}\), a completely positive trace-preserving map,  refers to the noise with \(p\) being the strength of the noise. For a fixed noise, \(\Lambda^{(p)}\), and a fixed unitary operator, \(U\),  the maximization is performed over the set of pure product states, denoted as \(S\), i.e., \(\{|\psi_1\rangle \otimes |\psi_2\rangle \} \in S \). Note that \(\Lambda^{(p)}\) can be global noise which acts on the entire system together or can be local noise, which is more natural to assume and in this work, we restrict ourselves to local noise, \(\Lambda_1^{(p_1)} \otimes \Lambda_2^{(p_2)}\).  Therefore, for a given local noisy channel acted on the product inputs, we have
\begin{eqnarray}
\mathcal{E}^{\Lambda} (U) = \max_{S} \mathcal{E} [U(\Lambda_1^{(p_1)}(|\psi_1\rangle \langle \psi_1 |) \otimes \Lambda_2^{(p_2)}(|\psi_2\rangle \langle \psi_2 |))U^\dagger].
\end{eqnarray}
where \(\rho_1=\Lambda_1^{(p_1)}(|\psi_1\rangle \langle \psi_1 |)\) and \(\rho_2=\Lambda_2^{(p_2)}(|\psi_2\rangle \langle \psi_2 |)\) are the output states after the action of local noise.
In a noiseless scenario, suppose the product state, \(|\psi_1^{opt}\rangle \otimes |\psi_2^{opt}\rangle \), is an optimal input for the action of a unitary $U$. This is an intriguing question of whether the same optimal state, under noise, also produces the maximal entanglement or not for the same operator. We will demonstrate that depending on the character of the noise, the optimal product input indeed differs from the one that is obtained for the noiseless case. This prior knowledge about the noise would help to decide which input state can be inserted in the circuit. 

The above formulation can easily be generalized to a multipartite domain. As shown in a recent work by some of us \cite{SamirAditi23}, since several kinds of separable states \cite{Dur01} exist in a multipartite domain,  several classes of multipartite entangling power emerge, depending on the set chosen for the optimization in Eq. (\ref{eq:entpower}). Here we can assume that some or all of the subsystems of inputs can be affected by noise, accordingly, the entangling power gets changed which we will address in the succeeding section. Moreover,  due to the action of the noisy channel(s),  the pure input states convert to mixed states, and hence the computation of the entangling power of a unitary operator requires quantification of entanglement for mixed states, which is, in general, difficult to compute, especially in a multipartite domain.




\section{Consequence of imperfections in unitaries on entangling power with Two-qubit inputs}
\label{sec:imperfectunitarytwoq}

We will now explore the effects of imperfections on the entangling power of quantum gates for a two-party scenario. In particular, we focus on $\mathcal{E}(U)$ for diagonal unitaries, and any generic two-qubit unitaries whose parameters are not fixed to the desired values. 


\subsection{Imperfection in diagonal unitaries}

Diagonal unitaries play an important role in the implementation of a quantum computer, especially in the measurement-based quantum computation \cite{Briegel2001, Briegel2009} and hence before considering arbitrary two-qubit gates, we focus on the noisy entangling power of diagonal unitaries with imperfections. Let us consider an arbitrary four-dimensional diagonal unitary operator, given by 
\begin{eqnarray}
    U_{d,4}^{4}=\text{diag} (e^{i\phi_{1}},e^{i\phi_{2}},e^{i\phi_{3}},e^{i\phi_{4}}),
    \label{eq:diag_4_4}    
\end{eqnarray}
with $\phi_{i} \in (0,2\pi)$  $(i=1,2,3,4)$. Here the superscript denotes the dimension in which the unitary is defined. In the subscript, $d$ is used to mean that the operators are diagonal and the numerical value ``4''  represents the number of nonvanishing $\phi_{i}$'s. Note that if we allow $\phi_{i}$'s to take negative values, the operators still remain unitary.

As discussed in Sec. \ref{sec:frame}, suppose we want to execute a diagonal unitary operator with a fixed set of $\{\phi_{i}\}_{i=1}^{4}$. Due to the imperfections in the device, instead of $\{\phi_{i}\}$s, they are chosen from  Gaussian distributions,  $G(\langle \phi_{i} \rangle,\sigma_{\langle \phi_{i} \rangle})$s, with mean \(\langle \phi_{i} \rangle\) and standard deviation, \(\sigma_{\langle \phi_{i} \rangle}\). We evaluate the corresponding  quenched average entangling power $\mathcal{E}_{\mathcal{G}}^{\text{avg}}(U_{d,4}^4(\{\langle \phi_{i} \rangle,\sigma_{\langle \phi_{i} \rangle}\}))$. We choose  negativity $N$, \cite{Vidal02} and generalized geometric measure (GGM), $\mathcal{G}$ \cite{aditi2010} to quantify entanglement (for definition, see Appendix $A$). The first one is chosen because it can be computed easily even in the presence of noise while the second one can be easily computed to quantify genuine multipartite entanglement for multipartite pure states. 

Before finding the quenched average entangling power, let us first note an important observation with respect to the input states. Specifically, we prove that the entangling power of the single-parameter arbitrary diagonal unitary operators, $U_{d,1}^{4} (\phi)=\text{diag}(1,1,1,e^{i\phi})$ 
(with \(\phi_4\equiv\phi \)) is attained at the same input states, so the inputs are independent of $\phi$. 


\textbf{Proposition I.} {\it The entanglement that can be created by an arbitrary single-parameter diagonal unitary operator, $U_{d,1}^{4} (\phi)$ always gets maximized when it acts on a product state of the form   $\frac{1}{\sqrt{2}}(\ket{0}+e^{i\xi_{1}}\ket{1})\otimes \frac{1}{\sqrt{2}} (\ket{0}+e^{i\xi_{2}}\ket{1})$ and the corresponding entangling power in terms of GGM and negativity are given by \(\min[\cos^2 \phi/4, \sin^2 \phi/4]\) and \(N=\frac{1}{16} (4 | -1+e^{i \phi }| +| 2 \sqrt{e^{i \phi } (1+e^{i \phi })^2}+4 e^{i \phi }| +| 4 e^{i \phi }- 2 \sqrt{e^{i \phi } (1+e^{i \phi})^2}| -8)\) respectively.}

\begin{proof}
Let us consider the diagonal unitary operator $U_{d,1}^{4} (\phi)$ which  acts on an arbitrary two-qubit product state, given by
\begin{eqnarray}
    \ket{\psi^{sep}}&=&\otimes_{k=1}^{2}\cos\theta_{k}\ket{0}+\sin\theta_{k}e^{i\xi_{k}} \ket{1},
    \label{eq:state_product}
\end{eqnarray}
with $0\leq \theta_{k} \leq \pi$ and $0\leq \xi_{k} \leq 2\pi$. Let us first elaborate on how we obtain the optimal state in the case of GGM. A similar procedure leads to the same input product state in the case of negativity. The entangling power of the resulting state is quantified by GGM as 
  \( \mathcal{E}_{\mathcal{G}}(U_{d,1}^{4} (\phi)) =\max_{\theta_{1},\theta_{2},\xi_{1},\xi_{2}} \mathcal{E}_{\mathcal{G}}(U_{d,1}^{4} (\otimes_{k=1}^{2}\cos\theta_{k}\ket{0}+\sin\theta_{k}e^{i\xi_{k}} \ket{1}))\).
The reduced density matrices required to calculate GGM, in this case, read as \begin{eqnarray}
\rho_{i}&=&
\left(
\begin{array}{cc}
 \cos^2\theta_{k} & e^{-i (\phi +\xi_k)} Q_{k}  \\
 e^{i\xi_{k}} R_{k}  & \sin^2\theta_{k}, \\
\end{array}
\right), (k=1,2)
\end{eqnarray}
where $Q_{1}=\sin \theta_{1} \cos \theta_1 (\sin^2\theta_{2}+e^{i \phi } \cos^2\theta_{2})$, and  $R_{1}=\sin \theta_{1} \cos \theta_{1} (\cos^2\theta_{2}+e^{i \phi} \sin^2\theta_{2})$. 
\(Q_2\) and \(R_2\) can be obtained by replacing \(\theta_1\) by \(\theta_2\) and vice-versa. 

Therefore,  $\mathcal{E}_{\mathcal{G}}(U_{d,1}^{4}(\phi))=\max[1-\max[\lambda_{1},\lambda_{2}]]$ with 
   \(\lambda_{1,2}= \frac{1}{8} (4\pm \sqrt{(8 \sin^2 2\theta_{1} \sin^2 2\theta_{2} \cos \phi - A})\), where \(A =\cos 4 (\theta_{1}-\theta_{2})+\cos 4 (\theta_{1}+\theta_{2})- 2 \cos 4 \theta_{1} - 2 \cos 4 \theta_{2}-14)\). 
For a fixed \(\phi\), maximization over \(\theta_k\)s and \(\xi_k\)s lead to the condition that $\theta_{1}=\theta_{2}=\frac{\pi}{4}$ $\forall \xi_{k}$. Hence the corresponding entangling power in terms of GGM simplifies as 
$\mathcal{E}_{\mathcal{G}}(U_{d,1}^{4}(\phi)) = \min[\cos^2 \phi/4, \sin^2 \phi/4]$. 

A similar exercise for negativity can show that the optimal product state is the same as that obtained for  GGM. By considering the optimal input, the entangling power in terms of negativity can be achieved as mentioned in the proposition (see Appendix B for details).


\end{proof}

\textbf{Remark 1.} Although we have shown the above proposition for arbitrary diagonal unitary operator with a single parameter,  we numerically find that for all diagonal unitaries, the optimal state is $\otimes_{k=1}^{2}\frac{1}{\sqrt{2}}(\ket{0}+e^{i\xi_{k}} \ket{1})$. The entangling power with GGM  for arbitrary diagonal unitary operators, \(U_{d,4}^4(\phi_1, \phi_2, \phi_3, \phi_4)\), reads as 
 \begin{eqnarray}    
 && \mathcal{E}_{\mathcal{G}} (U_{d,4}^4(\phi_1, \phi_2, \phi_3, \phi_4))=\frac{1}{8} e^{-i (\sum_{i=1}^{4}{\phi_{i}})}\nonumber\\
 &&(4 e^{i (\sum_{i=1}^{4}{\phi_{i}})}-2 \sqrt{e^{i (\sum_{i=1}^{4}{\phi_{i}})} \left(e^{i ({\phi_1}+{\phi_4})}+e^{i ({\phi_2}+{\phi_3})}\right)^2}),\nonumber\\
 \label{eq:powerGGMdiagonalarbitwo}
 \end{eqnarray}
 while in the case of  negativity, it takes the form  
 \begin{eqnarray}
 &&\mathcal{E}_{N}(U_{d,4}^4(\phi_1, \phi_2, \phi_3, \phi_4))=\frac{1}{8} \big(2 | e^{i ({\phi_2}+{\phi_3})}-e^{i ({\phi_1}+{\phi_4})}|\nonumber \\
 &&+| \sqrt{e^{i(\sum_{i=1}^{4} {\phi_i}) }(e^{i ({\phi_2}+{\phi_3})}+e^{i ({\phi_1}+{\phi_4})})^2}-2 e^{i(\sum_{i=1}^{4} {\phi_i})}|\nonumber\\
 &&+| \sqrt{e^{i (\sum_{i=1}^{4}{\phi_i})} \left(e^{i ({\phi_1}+{\phi_4})}+e^{i ({\phi_2}+{\phi_3})}\right)^2}+2 e^{i (\sum_{i=1}^{4}{\phi_i})}|.\nonumber\\
 \label{eq:powerNegdiagonalarbitwo}
 \end{eqnarray}

\subsubsection{Quenched average entangling Power}
\label{subsec:quench}

\begin{figure}
\includegraphics [width=\linewidth]{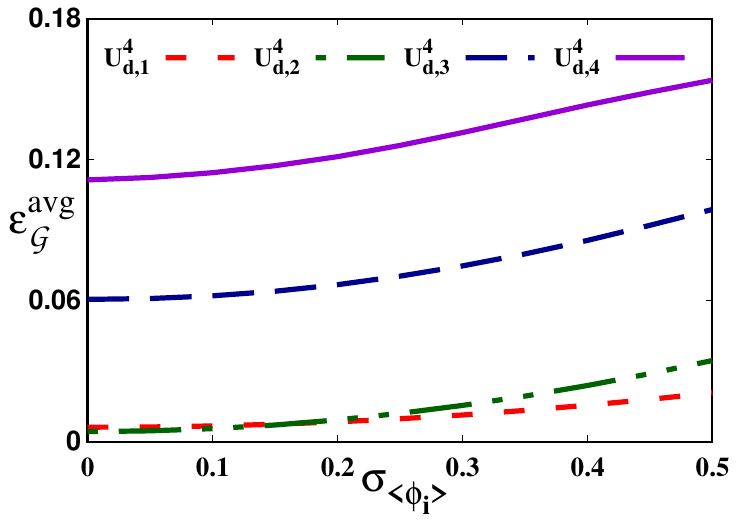}
\caption{{\it Constructive effects of disorder.} Quenched average entangling power in terms of GGM,  $\mathcal{E}_{\mathcal{G}}^{avg}$ (ordinate) of the diagonal unitaries against disorder strength, $\sigma_{\langle\phi_{i}\rangle}$ (abscissa). Different curves represent different unitaries, $U^{4}_{d,1}$, $U^{4}_{d,2}$, $U^{4}_{d,3}$ and $U^{4}_{d,4}$ with different means $\langle \phi_{i} \rangle$ .  For $U^{4}_{d,1},\langle\phi\rangle=\frac{\pi}{10}$ (red dashed line), for $U^{4}_{d,2},\langle\phi_1\rangle=\frac{\pi}{10},\langle\phi_2\rangle=\frac{\pi}{6},$(green dashed line), while  $\langle\phi_1\rangle=\frac{\pi}{10},\langle\phi_2\rangle=\frac{\pi}{6},\langle\phi_3\rangle=\frac{\pi}{4}$ is chosen for  $U^{4}_{d,3}$, (blue dashed line), and $\langle\phi_1\rangle=\frac{\pi}{10},\langle\phi_2\rangle=\frac{\pi}{6},\langle\phi_3\rangle=\frac{\pi}{4},\langle\phi_4\rangle=\frac{3\pi}{4}$ for $U^{4}_{d,4}$. In the presence of disorder, the quenched average GGM increases with the increase of disorder strength for all four unitaries, thereby demonstrating the positive impact of imperfections. Both axes are dimensionless.}
\label{fig:constructive}
\end{figure}

\begin{figure}
\includegraphics [width=\linewidth]{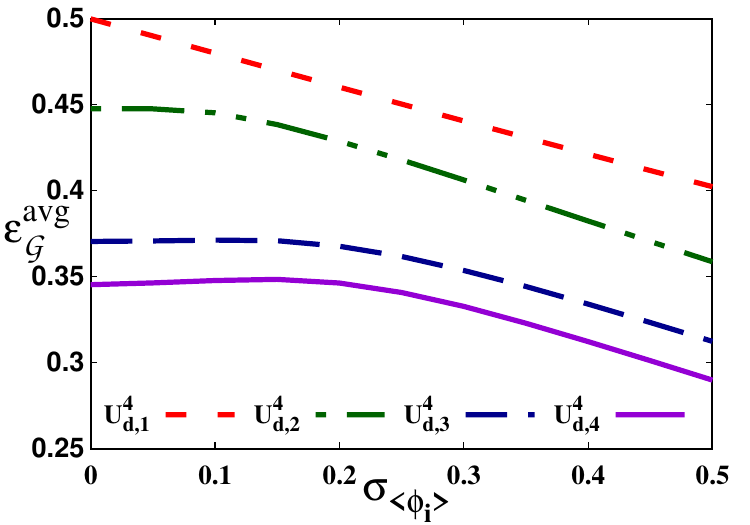}
\caption{ {\it Destructive impacts of disorder.} Quenched average entangling power with GGM,  $\mathcal{E}_{\mathcal{G}}^{avg}$ (vertical axis) against disorder, $\sigma_{\langle\phi_{i}\rangle}$ (horizontal axis). 
Curves are for $U^{4}_{d,1},\langle\phi\rangle={\pi}$ (red dashed line), for $U^{4}_{d,2},\langle\phi_1\rangle={\pi},\langle\phi_2\rangle=\frac{\pi}{15},$(green dashed line), for $U^{4}_{d,3},\langle\phi_1\rangle={\pi},\langle\phi_2\rangle=\frac{\pi}{10},\langle\phi_3\rangle=\frac{\pi}{15}$ (blue dashed line), and for $U^{4}_{d,4},\langle\phi_1\rangle={\pi},\langle\phi_2\rangle=\frac{\pi}{6},\langle\phi_3\rangle=\frac{\pi}{10},\langle\phi_4\rangle=\frac{\pi}{15}$. The average GGM decreases as one expects due to faulty device. The vertical and horizontal axes are dimensionless.}
\label{fig:destructive}
\end{figure}

Due to the above proposition and the numerical evidences we have,  in the quenched average entangling power, maximization over the set of product states is not required. 
With this simplification in hand, 
the expression for the quenched average entangling power for \sm{ \(U_{d,1}^{4} (\langle \phi \rangle, \sigma_{\langle \phi\rangle})\)} becomes 

\begin{eqnarray}
   \mathcal{E}_{\mathcal{G}}^{ avg} (U_{d,1}^{4} (\langle \phi \rangle, \sigma_{\langle \phi\rangle})) = &&\frac{1}{\sqrt{2\pi}} \int_{-\infty}^{\infty} \min[\cos^2 \frac{\phi}{4}, \sin^2  \frac{\phi}{4}] \times \nonumber \\
   &&\frac{1}{\sigma_{\langle \phi\rangle}} e^{-\frac{(\phi - \langle \phi\rangle)^2}{2 \sigma^{2}_{\langle \phi\rangle}}} d\phi.  
   \label{eq:entpoweravg2qubit}
\end{eqnarray}
The above quantity can increase or decrease with the strength of the disorder \(\sigma_{\langle \phi\rangle}\) as shown in Figs. \ref{fig:constructive} and \ref{fig:destructive}. In the presence of imperfections, one typically expects that the effects on the entangling power will be destructive although we will elaborate that this is not always true. 
\sud{After integration, Eq. (\ref{eq:entpoweravg2qubit}) becomes}
\begin{eqnarray}
\mathcal{E}_{\mathcal{G}}^{avg} (U_{d,1}^{4} (\langle \phi \rangle, \sigma_{\langle \phi\rangle}))&=&\frac{1}{4} e^{-\frac{\sigma_{\langle \phi\rangle} ^2}{8}-\frac{i \langle\phi\rangle}{2}} \nonumber \\
&&\left(2 e^{\frac{1}{8} \left(\sigma_{\langle \phi\rangle} ^2+4 i \langle\phi\rangle\right)}\pm e^{i \langle\phi\rangle} \pm 1\right),\,\,\nonumber\\
\label{eq:avg_ggm_sin}
\end{eqnarray}
\sud{where \(``+"\) and \(``-"\) correspond to  $\cos^{2}{\frac{\phi}{4}}$ and   $\sin^{2}{\frac{\phi}{4}}$ respectively in the minimization involved in Eq. (\ref{eq:entpoweravg2qubit}).}

\begin{figure}
\includegraphics [width=\linewidth]{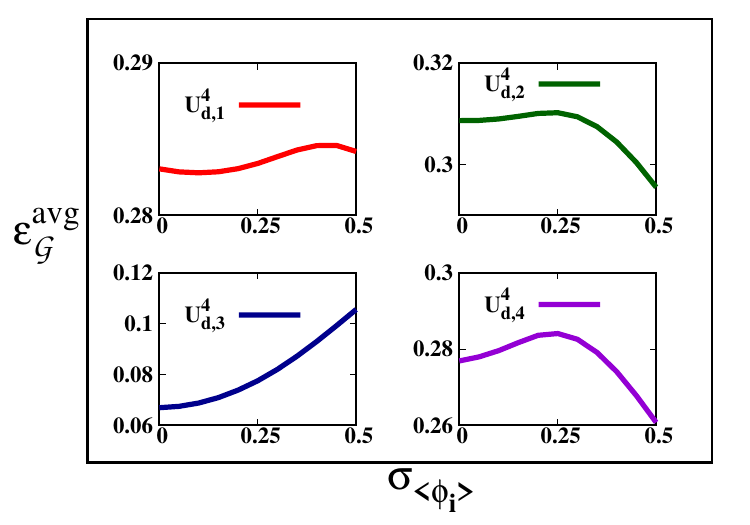}
\caption{{\it Nonmonotonic behavior due to disorder.} The quenched average entangling power in terms of GGM,  $\mathcal{E}_{\mathcal{G}}^{avg}$ (vertical axis) with respect to  disorder strength, $\sigma_{\langle\phi_{i}\rangle}$ (horizontal axis). For diagonal unitaries $U^{4}_{d,1}$, $U^{4}_{d,2}$, $U^{4}_{d,3}$ and $U^{4}_{d,4}$, the means are chosen respectively $\langle\phi\rangle=\frac{\pi}{1.3}$ (red solid line), $\langle\phi_1\rangle={\pi},\langle\phi_2\rangle=\frac{\pi}{4},$(green solid line), $\langle\phi_1\rangle={\pi},\langle\phi_2\rangle=\frac{\pi}{6},\langle\phi_3\rangle=\frac{\pi}{2}$ (blue solid line), and $\langle\phi_1\rangle={\pi},\langle\phi_2\rangle=\frac{\pi}{4},\langle\phi_3\rangle=\frac{\pi}{9},\langle\phi_4\rangle=\frac{\pi}{15}$.
The quenched average  GGMs show counter-intuitive behavior, both increase and decrease with the increase of disorder $\sigma_{\langle\phi_{i}\rangle}$. This is due to the competition of positive and negative influence of imperfections.  Both the axes are dimensionless.}
\label{fig:nonmonotonic}
\end{figure}

\begin{figure}
\includegraphics [width=\linewidth]{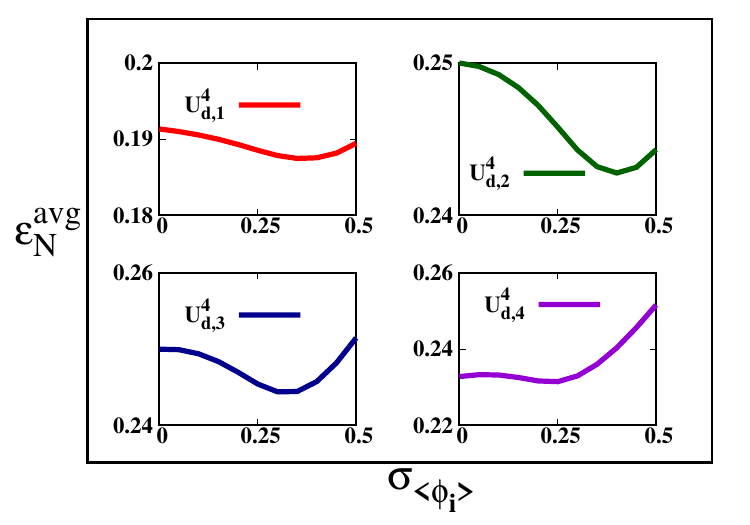}
\caption{Similar nonmonotonic trends of quenched average entangling power in terms of negativity,  $\mathcal{E}_{N}^{avg}$ (y-axis) with $\sigma_{\langle\phi_{i}\rangle}$ (x-axis) emerges as shown in Fig. \ref{fig:nonmonotonic}.  Curves represent the diagonal unitaries,  $U^{4}_{d,1}$ with $\langle\phi\rangle=\frac{\pi}{4}$ (red solid line),  $U^{4}_{d,2}$ with $\langle\phi_1\rangle=\frac{\pi}{2},\langle\phi_2\rangle=\frac{\pi}{6},$ (green solid line),  $U^{4}_{d,3}$ with $\langle\phi_1\rangle=\frac{\pi}{2},\langle\phi_2\rangle=\frac{\pi}{10},\langle\phi_3\rangle=\frac{\pi}{15}$ (blue solid line), and  $U^{4}_{d,4}$ with $\langle\phi_1\rangle=\frac{\pi}{2},\langle\phi_2\rangle=\frac{\pi}{6},\langle\phi_3\rangle=\frac{\pi}{8},\langle\phi_4\rangle=\frac{\pi}{10}$. Note that means are chosen to be different from Fig. \ref{fig:nonmonotonic}. Both the axes are dimensionless.}
\label{fig:nonmonotonicNeg}
\end{figure}
 Although an exact expressions for the entangling power of an arbitrary diagonal unitary operator, $U_{d,k}^{4}$ $(k=1,2,3)$  in Eqs. (\ref{eq:powerGGMdiagonalarbitwo}) and (\ref{eq:powerNegdiagonalarbitwo}) can be obtained, the compact form for the quenched average entangling power 
 cannot be analytically found when more than one $\phi_{i}$s are nonvanishing. In these situations, we numerically perform averaging over $10^{4}$ realizations which are fixed by the convergence of the quantity, and observe some interesting features which are noted below. The typical expectation is that any faults in the device can have adverse effects on the performance which in this case, is the entangling power of the gate.  Let us first show that this is indeed the case in some scenarios although the positive aftermath can also be observed. 
 
$1.$ {\it Destructive impact of disorder.}  Let us first consider the situations where the unfavorable effects of disorder are seen. In particular, fixing \(U_{d,1}^{4}(\phi)\) with \(\phi =\pi\), it is possible to find the input state from which a maximally entangled state can be created. On the other hand,  \sm{when the exact value cannot be tuned, we consider the mean  \(\langle\phi\rangle\) to be \(\pi\), and that} leads to the decrement in entanglement-generation (see Fig. \ref{fig:destructive}). This is a trivial situation because when entanglement generation by a given quantum gate reaches the maximum achievable value without disorder, the only possibility due to the introduction of disorder is the decrease in entanglement with the increase of disorder strength. In this situation, the nontrivial scenario emerges when the entangling power of a given unitary operator does not reach maximum but still decreases with the increase of disorder, as shown in Fig. \ref{fig:destructive} for \(U_{d,k}^4\) (\(k=2, 3, 4\)).

\(2.\) \emph{Constructive Impact of disorder.} 
The imperfection inducing the improvement of the performance of the device can always be counter-intuitive. We observe that it is possible to find unitary operators \( U_{d,k}^{k}(\phi_i)\) for all \(k=1,2,3,4\) and the set of \(\phi_i\)s  which have \(\mathcal{E}_{\mathcal{G}(N)} \neq 0\) in the perfect execution although with the imperfections, its entangling power gets enhanced, i.e.,  \(\mathcal{E}_{\mathcal{G}(N)} < \mathcal{E}^{avg}_{\mathcal{G}(N)}\). Precisely, \(\mathcal{E}^{avg}_{\mathcal{G}(N)}\) increases with the increase of the disorder strength, \(\sigma_{\langle\phi_i\rangle}\), 
as depicted in Fig. \ref{fig:constructive}, thereby leading to a nontrivial demonstration of positive consequence of disorder.  Note that one can also consider a trivial example here when all the $\phi_{i}$s are same, representing an identity operator, and hence $\mathcal{E}(U_{d,4}^{4}(\langle\phi_i\rangle = \langle\phi\rangle))=0$ although when all of them are chosen from $G(\langle\phi\rangle =0,\sigma_{\langle \phi\rangle})$ independently, 
they deviate from the identity operator, thereby capable of generating nonvanishing entanglement which increases with $\sigma_{\langle \phi\rangle}$.

\(3.\) {\it Nonmonotonic behavior with increasing disorder strength.} There can be some intuitions behind the increasing or decreasing trends of entangling power with the SD as presented before. However, we observe that \(\mathcal{E}^{avg}_{\mathcal{G}(N)} (U_{d,k}^4)\) (\(k=1,2,3,4\)) can show a nonmonotonic behavior with increase in \(\sigma_{\langle\phi_i\rangle}\).  Specifically, there exists a critical disorder strength below which it increases with \(\sigma_{\langle\phi_i\rangle}\) while it decreases above the critical value and vice-versa. The competition between constructive and destructive effects of disorder leads to these counter-intuitive nonmonotonic behaviors observed both in terms of GGM and negativity (see Figs. \ref{fig:nonmonotonic} and \ref{fig:nonmonotonicNeg}).

\subsection{Imperfection in two-qubit generic unitary}
\label{sec:imperfectgenericU}

\sud{It was proven that any arbitrary unitary operator acting on \(n\) qubits can be decomposed as well as parameterized via a single- and two-qubit unitary operators \cite{Khaneja2001, PhysRevA.52.3457}. The explicit construction is performed by using the Cartan decomposition of a semi-simple Lie group \cite{Khaneja2001}.} 
\sud{Specifically,  in a four-dimensional complex Hilbert space, it can be represented as \cite{Farrokh2004}
\begin{eqnarray}
\nonumber U_{\text{gen}}^{4} &=& \exp \bigg[i \bigg(\sum_{i,j} a_{ij} \sigma_i \otimes \sigma_j 
\\&+& \sum_{i} b_{i} (\sigma_{i} \otimes I) 
+ \sum_{j} c_{j} (I \otimes \sigma_{j}) \bigg) \bigg]. \nonumber\\
    \label{eq:4_gen_u}  
\end{eqnarray}
with real coefficients \(a_{ij}\), \(b_{i}\), \(c_{i}\).  Further, it can be rewritten as
\begin{eqnarray}
U_{\text{gen}}^{4}= (A_{1}\otimes A_{2}). {U_{NL}^{4}}.(A_{3}\otimes A_{4}),
\end{eqnarray}
where \(A_{j}\in \textbf{SU}(2)\)  and 
\begin{eqnarray}
     U_{NL}^{4}= \exp [-i \sum_{i=1,2,3} (J_i {\sigma_i}\otimes {\sigma_i})].
    \label{eq:4_gen}
\end{eqnarray}
Here $\sigma_{i}$s ($i=1,2,3$) are the Pauli matrices and $J_{i}\in \mathbb{R}$, $\forall i=1,2,3$. Since local unitary operators cannot change the entanglement, we only focus on the entanglement capability of $U^4_{NL}$. Suppose all $J_i$s are equal to \(J\) \(\forall i\). Moreover, we assume that it cannot be tuned appropriately, and hence it is chosen from a Gaussian distribution with mean \(\langle J \rangle\) and SD \(\sigma_{\langle J\rangle}\). In this case, the input state optimizing the entangling power of \(U_{NL}^4\) for all realizations can be proven in the following proposition. }

\begin{figure}
\includegraphics [width=\linewidth]{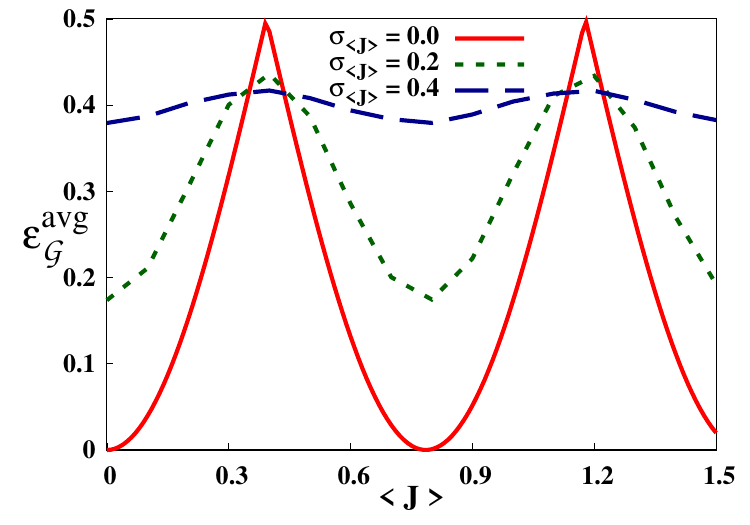}
\caption{Quenched average entangling power with GGM,  $\mathcal{E}_{\mathcal{G}}^{avg}(U_{NL}(\{\langle J \rangle, \sigma_{\langle J\rangle}\}))$ (vertical axis) for a specific class of  generic two-qubit unitaries, given in 
 Eq. (\ref{eq:4_gen}) with \(J_1 = J_2 = J_3=J\), against the mean, \(\langle J\rangle\) (horizontal axis). Various curves correspond to various disorder strength,  $\sigma_{\langle J \rangle}=0.0$ (solid red line), and $\sigma_{\langle J \rangle}=0.2$ (dashed green line), $\sigma_{\langle J \rangle}=0.4$ (dashed blue line). Both the vertical and horizontal axes are dimensionless.}
\label{fig:gen_allJeq}
\end{figure}

\begin{figure}
\includegraphics [width=\linewidth]{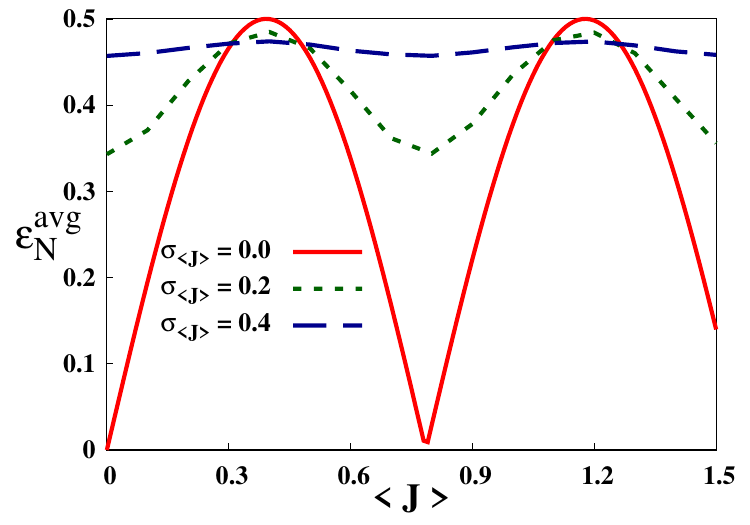}
\caption{Quenched average entangling power in terms of negativity,  \(\mathcal{E}_{N}^{avg}(U_{NL}^4(\{\langle J \rangle, \sigma_{\langle J\rangle}\}))\) (ordinate) against $\langle J \rangle$ (abscissa) for generic unitary, $U_{NL}^{4}\sh{(J)}$ . All other specifications are the same as in Fig. \ref{fig:gen_allJeq}.
Both the ordinate and abscissa are dimensionless.}
\label{fig:gen_allequal_neg}
\end{figure}

\begin{figure}
\includegraphics [width=\linewidth]{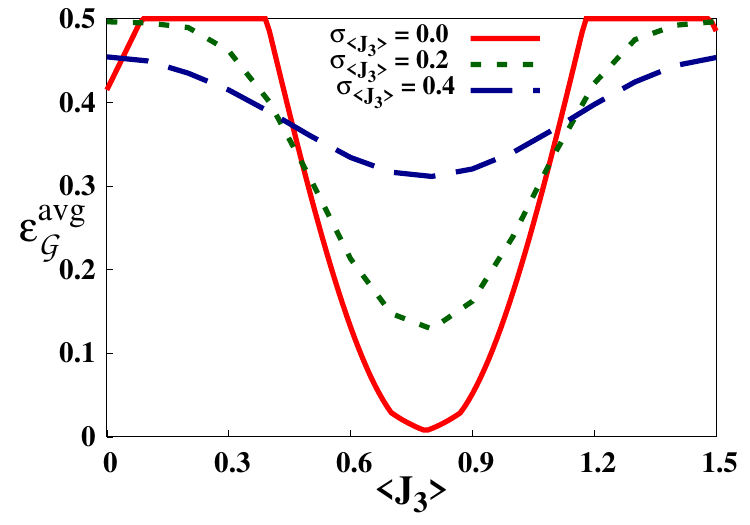}
\caption{Quenched average entangling power in terms of GGM,  $\mathcal{E}_{\mathcal{G}}^{avg}(U_{NL}^4(J_1,\{\langle J_2\rangle=\langle J_3\rangle,\sigma_{\langle J_3\rangle}\}))$ (vertical axis) against $\langle J_{3} \rangle$ (horizontal axis) for generic two-qubit unitary, \(U_{NL}^{4}(J_1,\{ J_2= J_3\}\})\) having $J_{1}=0.7$ and $\langle J_{2}\rangle = \langle J_{3} \rangle$. All other specifications are the same as in Fig. \ref{fig:gen_allJeq}. The above pattern of \(\mathcal{E}_\mathcal{G}^{avg}\) displays that this oscillation-free behavior with high disorder strength is universal for arbitrary two-qubit unitaries. 
Both the axes are dimensionless.}
\label{fig:gen_allJunequal}
\end{figure}

\textbf{Proposition II.} {\it When \sh{\(U_{NL}^{4}(J)\)} acts on the product states, the maximization in \(\sh{\mathcal{E}(U_{NL}^{4}(J))}\) is achieved with the input state} being either $\ket{01}$ or $\ket{10}$ and the corresponding entangling power in terms of GGM and negativity can be computed.

\begin{proof}
    When $\sh{U_{NL}^{4} (J)}$ acts on any arbitrary product state, $\ket{\psi^{sep}}=\otimes_{k=1}^{2}(\cos{\theta_{k}}\ket{0}+\sin{\theta_{k}}e^{i\xi_{k}}\ket{1})$, the resulting state takes the form
\sm{$\ket{\psi}=\sum_{i,j=0}^{1}a_{ij}\ket{ij}$},
Here for a single realization, $a_{00}=e^{-i J} \cos \theta _1 \cos \theta _2$, $a_{01}=\big(\frac{e^{-i  J}}{2}-\frac{e^{i3J}}{2}\big) e^{i\xi_1} \sin \theta_1 \cos \theta_2 + \big(\frac{e^{-i  J}}{2}+\frac{e^{i3J}}{2}\big) e^{i\xi_2} \sin \theta_2 \cos \theta_1$, $a_{10}=\big(\frac{e^{-i  J}}{2}+\frac{e^{i3J}}{2}\big) e^{i\xi_1} \sin \theta_1 \cos \theta_2 + \big(\frac{e^{-i  J}}{2}-\frac{e^{i3J}}{2}\big) e^{i\xi_2} \sin \theta_2 \cos \theta_1$, and $a_{11}=\sin \theta _1 \sin \theta _2 e^{i(\xi _1+ \xi _2- J)}$. 
The eigenvalues corresponding to the reduced density matrices of the resulting state for a single realization are 
\begin{eqnarray}
\mathcal{\lambda}_{1,2}&=&\frac{1}{2}\pm\frac{1}{4} e^{-4 i J} \sqrt{\left(-1+e^{8 i J}\right)^2 \sin ^4\left(\theta _1-\theta _2\right)+4 e^{8 i J}},\nonumber\\
\label{eq:gen_eigen}
\end{eqnarray}
which lead to the entangling power with GGM as $\sh{\mathcal{E}_{\mathcal{G}}(U_{NL}^{4} (J)})=\max[1-\max[\lambda_{1},\lambda_{2}]]$ involving maximization of $\theta_{k}$s and \(\xi_k\)s $(k=1,2)$. It can be found that
$\sh{\mathcal{E}_{\mathcal{G}}(U_{NL}^{4}(J)})=\min [\cos^2 2J, \sin^2 2J]$ which is obtained when \(\theta_1 =0 \,\text{or}\, \pi/2\) and \(\theta_2 = \pi/2\, \text{or}\, 0\) for all values of \(\xi_k\)s, i.e., the input state is either of the form \(|01\rangle\) or \(|10\rangle\). \sud{It is important to note here that   the input state optimizing \(\mathcal{E}_\mathcal{G}(U_{NL}^4(\{J\}))\) does not depend on \(J\) and hence it remains same for all realizations. Therefore, we have 
\begin{eqnarray}
    \mathcal{E}_{\mathcal{G}}^{avg}(U_{NL}^{4}(\langle J \rangle,\sigma_{\langle J\rangle}))=\nonumber\\
    \frac{1}{\sqrt{2\pi}}\int \mathcal{E}_{\mathcal{G}}(U_{NL}^4(\{J\}))\frac{1}{\sigma_{\langle J\rangle}} e^{-\frac{(J-\langle J \rangle)}{2\sigma^2_{\langle J \rangle}}} dJ.\nonumber
\end{eqnarray}
}A similar analysis for negativity can also be performed. \sud{Hence, the proof.}
\end{proof}

\textbf{Remark 2.} Diagonalizing \(U_{NL}^{4}\sh{(J)}\),  we obtain
the eigenvalues as \(\{e^{-i J}, e^{-i J}, e^{-i J}, e^{i 3J}\}\) and the corresponding eigenvectors  as \(\{|00\rangle, (|01\rangle+|10\rangle)/\sqrt{2}, (|01\rangle-|10\rangle)/\sqrt{2}, |11\rangle\}\) for a single realization and, importantly the eigenvectors are independent of \(J\). This possibly indicates the reason behind the similarity in the expressions obtained for the entangling power of the diagonal unitary operators with a single nonvanishing \(\phi\) and \(U_{NL}^{4}\sh{(J)}\).

Let us now estimate the entangling power when all the \(J_i\) are unequal. Unlike $U_{NL}^{4} \sh{(J)}$ proved in Proposition II, 
we numerically find that the input state $\ket{\psi^{sep}}$ that maximizes the entangling power upon the action of $U_{NL}^{4}\sh{(\{J_{1},J_2,J_3\})}$  has nonvanishing $\theta_{k}$s and $\xi_{k}$s. 
Notice further that the individual subsystems of the optimal input state for \(U_{NL}^{4}\sh{(J)}\) do not possess any quantum coherence with respect to the computational basis while for arbitrary \(J\) values, the reduced density matrices of the optimal input state possesses nonvanishing coherence \cite{coherenceRMP}.

{\it Oscillation-free entangling power with imperfections.} When there is no disorder in \(J_i\) and all of them are equal to \(J\),  the entangling power \(\mathcal{E}^{avg}_{\mathcal{G}(N)} (U_{NL}^{4}\sh{(\langle J\rangle,\sigma_{\langle J\rangle})})\) oscillates with the increase of \(J\). When \(J\) is chosen from \(G(\langle J\rangle, \sigma_{\langle J\rangle})\), the smoother behavior of \(\mathcal{E}^{avg}_{\mathcal{G} (N)} (U_{NL}^{4}\sh{(\langle J \rangle,\sigma_{\langle J \rangle})})\) emerges (see Fig. \ref{fig:gen_allJeq}) although the maximum power in the disorder case is lower than that obtained in the disorder-free case. 
When the standard deviation \(\sigma_{\langle J\rangle}\) of the disorder covers
at least one period of oscillation in \(\langle J\rangle\), the power becomes more or less independent of the mean, \(\langle J \rangle\). The oscillation-free 
\(\mathcal{E}^{avg}_{\mathcal{G} (N)} (U_{NL}^{4}\sh{(\langle J \rangle,\sigma_{\langle J \rangle})})\) implies that there \sm{exists} a range of \(\langle J\rangle\) where clearly, disorder enhances entangling  capability of \(U_{NL}^{4}\sh{(\langle J \rangle,\sigma_{\langle J\rangle})}\). Therefore, 
such an oscillation-free entangling power of \(U_{NL}^{4}\sh{(\langle J \rangle,\sigma_{\langle J\rangle})}\) obtained in the presence of disorder is another definite positive impact of imperfections. 

When all the \(J_i\) are unequal, similar behavior of \(\mathcal{E}^{avg}_{\mathcal{G}(N)} (U_{NL}^{4}\sh{(\langle J_i \rangle, \sigma_{\langle J_i \rangle})})\) can also be observed when \(\sigma_{\langle J_i \rangle}\)s are nonvanishing. In the perfect case,  the competition between \(J_1\), \(J_2\) and \(J_3\) decides the period of the oscillation in the entangling power of \sh{\(U_{NL}^{4}\sh{(\langle J_{i} \rangle,\sigma_{\langle J_i\rangle})}\)}. Again, in the presence of disorder, the depth of oscillations in quenched average entangling power decreases with the increase of the disorder strength, as shown in Fig. \ref{fig:gen_allJunequal}.

\section{Impact of noisy inputs on entangling power}
\label{sec:noisyentpower2qu}
 
We have already shown that when the desired quantum gates are not implemented perfectly, the optimal input states may not remain unchanged for arbitrary two-qubit unitary operators. However, some exceptions are also found for the classes of unitary operators as shown in Propositions I and II. \sm{Let us now examine whether the maximum entangling power is achieved with an input product state different from the one obtained in the noiseless case when the unitary operator and the noise model affecting the input states are fixed.} It implies that according to the noise model,  the initial state has to be changed to maximize the resource production via quantum gates. For inspection, we consider three paradigmatic noise models, amplitude damping (ADC), phase damping (PDC), and depolarizing (DPC) channels. \sud{The Kraus operators for ADC, PDC and DPC are given, respectively, by
\begin{eqnarray}
\nonumber &&K^{ADC}_0=
\begin{pmatrix}
1&0\\0&\sqrt{1-p}
\end{pmatrix}, \hspace{0.2cm}
K^{ADC}_1=
\begin{pmatrix}
0&\sqrt{p}\\0&0
\end{pmatrix},\\ \nonumber
&&K_0^{PDC}=\sqrt{1-\frac{p}{2}} \mathbb{I}, \hspace{0.2cm}K_1^{PDC}=\sqrt{\frac{p}{2}}\sigma_z, \text{and} \\ 
&&K_0^{DPC}=\sqrt{1-\frac{3p}{4}}\mathbb{I},\hspace{0.2cm} K_i^{DPC}=\sqrt{\frac{p}{4}}\sigma_i, \forall i\in 1,2,3,\nonumber\\
 \label{eq:kraus_adc}
\end{eqnarray}
where $0\leq p \leq1$ and superscripts of the Kraus operators represent the channels.

It is well understood that a unitary operator \( U \) generally possesses a greater entangling power without  noise than that in the presence of noise. It is clear if we assume that the optimal input state for maximizing entanglement remains the same in both scenarios. Mathematically, if the entangling power of \( U \) is denoted by \( \mathcal{E}(U) \), and the quantity under local noise, parameterized by \( p \), given by \( \mathcal{E}^{\Lambda(p)}(U) \), the following relation holds:
\begin{eqnarray}
   && \mathcal{E}(U) \geq \mathcal{E}^{\Lambda(p)}(U)\nonumber \\
   &&=\max_{S} \mathcal{E}[U(\Lambda^{(p)}(|\psi_1\rangle \langle \psi_1 | \otimes |\psi_2\rangle \langle \psi_2 |)U^\dagger)] \nonumber \\
   &&=\max_{S}\mathcal{E}[U(\rho_1^{(p)}\otimes\rho_2^{(p)})U^\dagger],
\end{eqnarray} 
where the maximization is performed over the set of separable states, \(S\). Interestingly, the optimal product state that maximizes entanglement without noise (left-hand side of the first line) differs from the optimal product state in the presence of noise (right-hand side of the first line) although the degrading nature of entangling power of a fixed  unitary operator remains true under the influence of noise. However, this observation may lead to a change in the hierarchy among the unitaries as will be shown later.}

  {\it Amplitude damping channel is special for diagonal unitaries.} 
Suppose one of the parties of the initial two-qubit product state gets affected by noise and an entangling diagonal unitary acts on the noisy input state. \sm{Here}, we are interested in analyzing the entanglement content of the resulting state. It is seen that when $U_{d,k}^{4}$ (\(k=1,2,3,4\)) acts on the noisy input state, the entanglement content of the output state decreases with the increment of the noise strength for all the three channels like ADC, PDC, and DPC although the optimal state, $\frac{1}{\sqrt{2}}(\ket{0}+e^{i\xi_{1}}\ket{1})\otimes \frac{1}{\sqrt{2}}(\ket{0}+e^{i\xi_{2}}\ket{1})$ in Proposition I remains the same for the PDC and DPC channels. However, the optimal state does not remain the same in case of ADC. It implies that if the noise is apriori known to be amplitude damping and the strength of the noise is also known, the maximum resource production demands the corresponding modification of the initial state.   For example,  for the ADC channel with \(p=0.4\) which acts on one of the parties, the entangling power, \(\mathcal{E}_{N} (U_{d,1}^4)\) with \(\phi = \pi/4\)  is maximum with the input state, \(\otimes_{k=1}^2( \cos \theta_k \ket{0} + e^{i\xi_{k}}  \sin \theta_k \ket{1})\) where \(\theta_1 =0.69\), \(\theta_2 = 3 \pi/4\), \(\xi_1= 2.16\) and \(\xi_2 = 2.18\). 

\begin{figure}
\includegraphics [width=\linewidth]{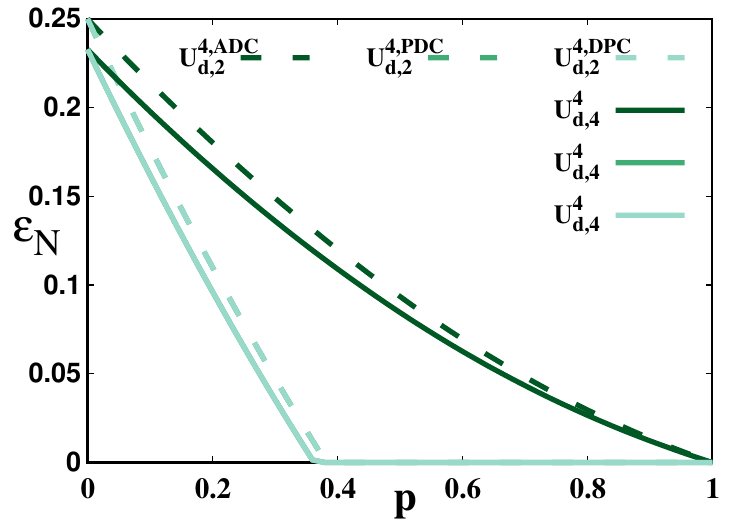}
\caption{Noisy entangling power ( with negativity), $\mathcal{E}_N$ (y-axis)  against local noise parameter, $p$ (x-axis) corresponding to diagonal unitaries, $U_{d,2}^4$ and $U_{d,4}^4$. Parameter values are fixed to the ones used in Fig. \ref{fig:nonmonotonicNeg} for disorder-free cases. The dashed line corresponds to $U_{d,2}^4$ while the solid line is for $U_{d,4}^4$. ADC, PDC, and DPC  (sequentially colored from dark green to light green) have impacts on entangling powers. Note that the entangling powers coincide when PDC and DPC act on input states. Both the axes are dimensionless.}
\label{fig:neg_noise_diagonal}
\end{figure}

\sm{It is interesting to note that}, even when both parties get affected by the noise, it does not have any extra impact on the optimal inputs, i.e., the optimal state still remains the same for the PDC as well as DPC and changes for ADC. However, the rate of decrease in the case of ADC is typically less than that of the PDC and DPC.  The decreasing behavior of \(\mathcal{E}_{N} (U_{d,k}^4)\) (\(k=2,4\)) with \(p\) for a fixed set of \(\phi_i\)s are shown in Fig. \ref{fig:neg_noise_diagonal}.  
\sud{Therefore, our numerical simulations suggest that the entangling powers of the diagonal unitary operators typically follow a specific ordering under noisy input states, given by}
\sud{
\begin{eqnarray} 
\mathcal{E}_{N}^{\text{PDC}} (U_{d,k}^4) =
\mathcal{E}_{N}^{\text{DPC}}(U_{d,k}^4) < \mathcal{E}_{N}^{\text{ADC}}(U_{d,k}^4),
\, k=1, \ldots 4.
\end{eqnarray}}
This illustrates that the effects of the ADC are minimal on the entangling power of diagonal unitary operators compared to the other channels.

\begin{figure*}
    
\includegraphics [width=0.85\linewidth, height=10.5cm]{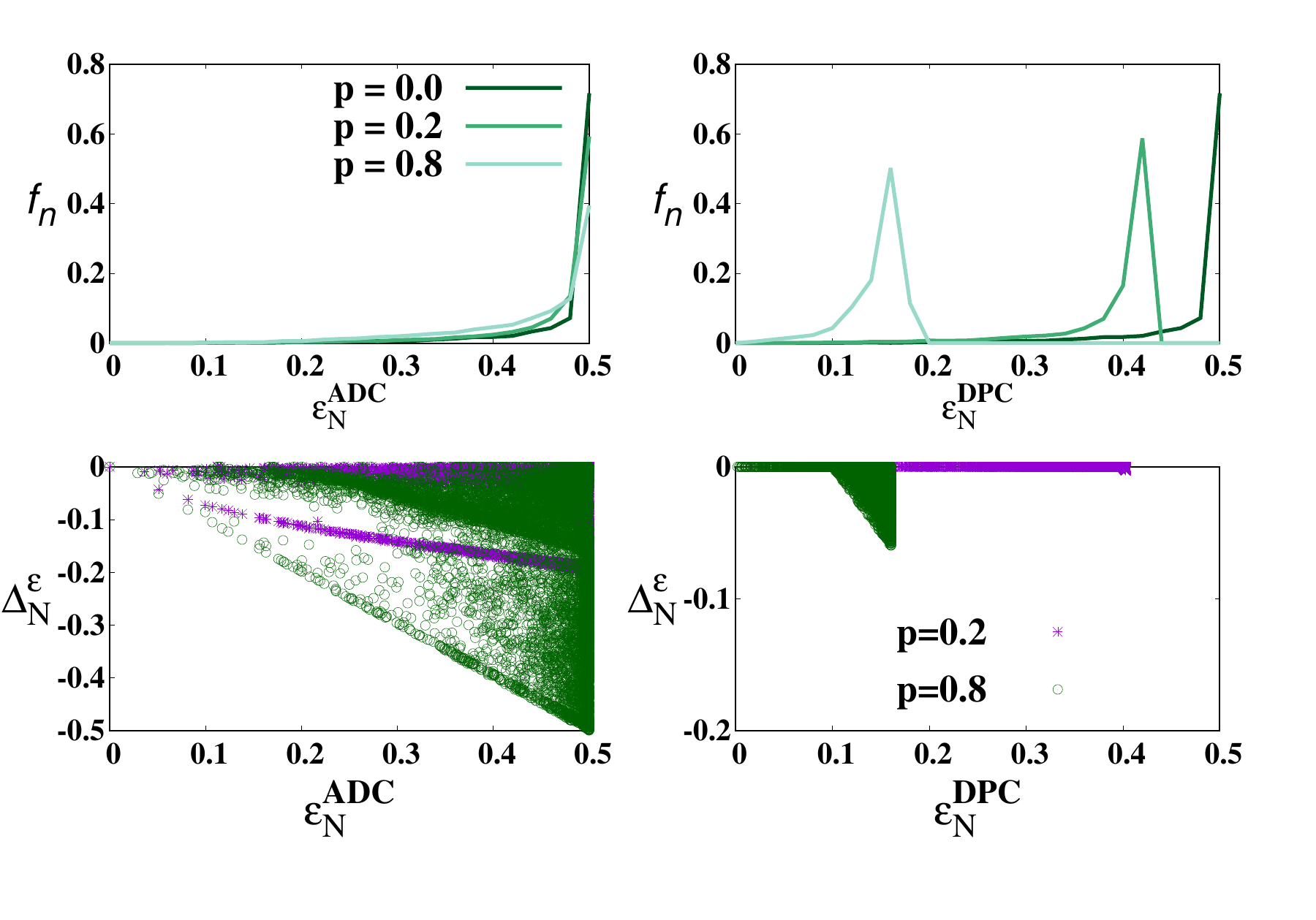}
\caption{Normalized frequency distribution of noisy entangling powers, $f_n$ and error, $\Delta^{\mathcal{E}}_{N}$  (ordinate) vs negativity,$\mathcal{E}^{ADC}_{N}$ or $\mathcal{E}^{DPC}_{N}$ (abscissa)  (for two channels ADC, DPC). (Upper panel) Different \(f_n\) curves are for noise parameters $p=0.0$ (noiseless case), $p=0.2$, and $p=0.8$ (sequentially colored). (Lower panel) The errors $\Delta^{\mathcal{E}}_{N}$ are with $p=0.2$ (violet star) and $p=0.8$ (green circles) . \(10^4\) number of unitaries  $U_{NL}^{4}\sh{(\{J_{i}\})}$ are randomly generated from uniform distribution. Both the axes are dimensionless.}
\label{fig:amp_n_1_u4gen}
\end{figure*}

{\it No effects of noise on entanglement generating power for special class of unitaries.} Let us now manifest that for a special class of arbitrary unitary operators, \sh{$U_{NL}^{4}(J)$, } with \(J_1= J_2 = J_3 =J\) in Eq. (\ref{eq:4_gen}), \sm{there exist some noise models affecting the inputs but not altering the entanglement content of the output states.  Specifically, we have the following proposition.}

\textbf{Proposition III.}  {\it When local  PDC acts on one of the parties or both the parties of the input bipartite product state (\(|01\rangle \langle 01|\)) which maximizes the entangling power of $U_{NL}^{4}\sh{(J)}$, the resulting state obtained after the action of $U_{NL}^{4}\sh{(J)}$, and its corresponding entanglement are independent of the noise parameter. The \sm{optimal} input state also remains the same when ADC acts on the first party of the bipartite inputs.  }

\begin{proof}
    As mentioned in Proposition II, the optimal input state that maximizes \(\mathcal{E}_{N}(U_{NL}^{4}\sh{(J)})\) with or without local amplitude damping noise being active on the first party is  \(|01\rangle\). \sm{It is also the same optimal input  when phase damping noise acts on one or both the parties. }\sud{ Let us consider the initial  density matrix, \(\rho_{in}=|01\rangle\langle01|\). After the action of local ADC on the first party, the noisy input state becomes \(\rho^{p}_{noisy}=\sum_{i=0}^{1}(K_{i}^{ADC}\otimes I)\rho_{in}(K_{i}^{ADC}\otimes I)^{\dagger}\). The final output state after the action of unitary, \(U_{NL}^{4}(J)\) is given by \(\rho_{out}=U_{NL}^{4}(J)\rho^p_{noisy}(U_{N }^{4}(J))^{\dagger}\), which turns out to be 
    \begin{eqnarray}
        \rho_{out}=\left(
\begin{array}{cccc}
 0 & 0 & 0 & 0 \\
 0 & L^2 & L T & 0 \\
 0 & L T & T^2 & 0 \\
 0 & 0 & 0 & 0 \\
\end{array}
\right),
    \end{eqnarray} 
where $L=\frac{e^{-i J}}{2}+\frac{1}{2} e^{3 i J}$, and $T=\frac{e^{-i J}}{2}-\frac{1}{2} e^{3 i J}$.} Interestingly, now if PDC acts on one  or both the parties of the same input \(|01\rangle\),   the resulting state after the action of the same unitary operator comes out to be exactly the same as $\rho_{out}$ obtained in the ADC case. 

 As it is clear from the expression, The output state $\rho_{out}$ is independent of the noise parameter \(p\)  and, therefore, the entangling power  of \(U_{NL}^{4}\sh{(J)}\) in the presence of ADC and PDC coincide with the noiseless case which can be calculated  as
    \begin{eqnarray}
    &&\mathcal{E}^{ADC,PDC}_{N}(U^{4}_{NL}\sh{(J)})\nonumber\\&&=\frac{1}{8} \big(| -1+e^{4 i J}| ^2+| 1+e^{4 i J}| ^2+2 | -1+e^{8 i J}| -4\big).\nonumber\\
\end{eqnarray}
Hence, the proof.
\end{proof}

When both the parties or one of the parties of the initial product state is affected by the local  DPC channel,  the input state that maximizes the entangling power of \(U^{4}_{NL}\sh{(J)}\) is again either \(|01\rangle\) or \(|10\rangle\). Now the entangling power depends on the noise parameter and in case of a single party  being disturbed, it can be given as 
   $\mathcal{E}^{DPC}_{N}(U^{4}_{NL}\sh{(J)})= \frac{1}{16} (| (-1+e^{4 i J})^2 (p-2)| +| (1+e^{4 i J})^2 (p-2)| +| 2 e^{5 i J} p+S| +| 2 e^{5 i J} p-S| -8)$, where $S=\sqrt{2} \sqrt{-e^{10 i J} (\cos (8 J) (p-2)^2+p (4-3 p)-4)}$. On the other hand, when local noise is in action on both parties of the input state, 
 
 it takes the form 
\(\mathcal{E}^{DPC}_{N}(U^{4}_{NL}\sh{(J)})=
\frac{1}{8} (| \sqrt{v}-e^{4 i j} (p-2) p| +| e^{4 i J} (p-2) p+\sqrt{v}| +| (e^{4 i J} (p-1)-1) (-p+e^{4 i J}+1)| +| (e^{4 i J} (p-1)+1) (p+e^{4 i J}-1)| -4)\),
 where $v=-(-1+e^{8 i J})^2 (p-1)^2$.
The expressions are valid under the assumption that the initial state is either \(|01\rangle\) or \(|10\rangle\) and the entangling power is nonvanishing. 

\begin{figure}
\includegraphics [width=\linewidth]{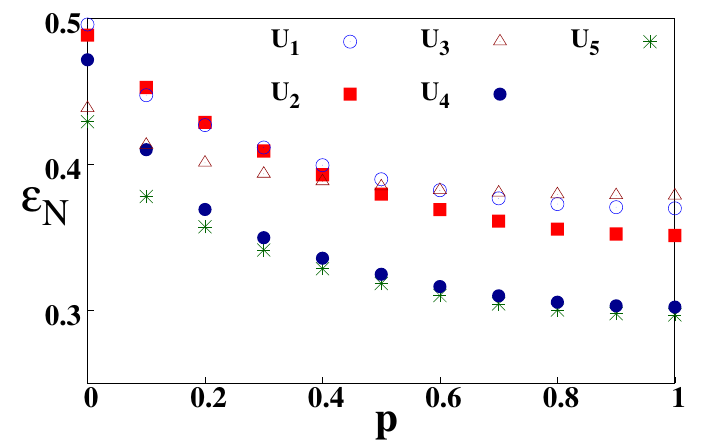}
\caption{{\it Change in hierarchy among unitaries.} Entangling power, $\mathcal{E}_N$ (vertical axis)  against noise parameter, $p$, in the phase damping channel (horizontal axis). The two-qubit unitaries are Haar uniformly generated and then they act on the set of product states in which both the parties are disturbed by the local dephasing noise. \sud{Squares, circles (solid and hollow), triangles, and stars represent five different Haar uniformly generated unitary matrices, \(U_i\) in the four-dimensional space (their forms are given in Appendix. \ref{app:mat_hhar}).} Note, interestingly, that the fall-off rates of entangling power in the presence of noise is different for different unitary operators. Hence, the ordering between \(U_1\) and \(U_2\), \(\mathcal{E}(U_1) > \mathcal{E}(U_2)\) with \(p=0\) may alter with the increase of \(p\), i.e., \(\mathcal{E}(U_1) < \mathcal{E}(U_2)\) may happen for some \(p>0\).  Both the axes are dimensionless.}
\label{fig:haar_deph}
\end{figure}

{\it Influence of noise on \sm{entangling power of}  arbitrary two-qubit unitary operators.} Let us now consider an arbitrary two-qubit unitary operator, \(U_{NL}^{4}\sh{(\{J_i\})}\). To scrutinize the role of noise on the entangling power,  we generate $10^4$ numbers of  \(U_{NL}^{4}\sh{(\{J_i\})}\) in which \(\{J_i\}\)s are chosen randomly from the uniform distribution over [$0,\frac{\pi}{2}$]. For each of the unitary operators and a given noise model with a specific noise strength \(p\), we compute the entangling power of that unitary operator after optimizing over the set of product states where all of the states are disturbed by the fixed local noise with noise strength, \(p\). 

To analyze the situation  carefully, we study the normalized frequency distribution,
\begin{eqnarray}
\sud{ f_{n}= \frac{\#\{U_{NL}^{4}(\{J_i\}): \mathcal{E}_{N}(U_{NL}^{4}(\{J_i\})) \in \mathcal{I}_N\}}{\text{Total number of operators simulated($=10^4$)}},
 \label{eq:Normal_freq}}
\end{eqnarray}
of the entangling power, $\mathcal{E}_{N}$. Here, we divide the range of entangling power, $[0,0.5]$, into 25 intervals of equal length and $\mathcal{I}_N$ refers to the N'th one of them.

Among \(10^4\) simulated unitary operators, we observe that in the noiseless situation, most of the unitaries are capable of creating highly entangled states; the mean of the distribution is obtained to be \(0.47507\) while the SD is \(0.05908\). With the introduction of noise, the capability of creating \sm{highly entangled states by the same unitary operator gets reduced and it becomes evident from the decrease of mean. \sud{Thus, increasing values of noise strength $p$ make $f_{n}$ to decrease which indicates that the results critically depend on the noise strength.} The reduction is more prominent for the DPC (see Fig. \ref{fig:amp_n_1_u4gen}) compared to other channels. For example, in the top two figures in Fig. \ref{fig:amp_n_1_u4gen},  for $p=0.8$,
in the case of ADC,  the mean and the SD are respectively given by \(0.438\) and \(0.08196\) (similar behavior emerges for the PDC) while in the case of DPC, they are \(0.146\) and \(0.0286\)} which means the entangling power gets reduced with the increasing value of noise strength. The figures also show that  $p=0.2$ case sits in between.

Let us now ask the following question: Instead of taking the optimal input state for the underlying  noise model, if one computes the entanglement of the output state under the action of the given unitary operator and the same noise on the optimal input state for the noiseless case, \sm{how much of difference we encounter?}  
To quantify it, we introduce a quantity, called error in the entangling power which is defined  as 
\begin{eqnarray}
\Delta^{N}_{\mathcal{E}}&=& \mathcal{E}^{\Lambda(p)}_{N}(U)-{E}^{\Lambda(p)}_{N}(U).
    \label{eq:error_def}
\end{eqnarray}
Here 
$\mathcal{E}^{\Lambda(p)}_{N}(U) = \mathcal{E}_{N} (U(\Lambda_1 (|\psi'_1\rangle \langle \psi'_1|) \otimes |\psi'_2\rangle \langle \psi'_2|)U^\dagger) $ where \(|\psi'_1\rangle \otimes |\psi'_2\rangle\) is the optimal product state for the noise model under consideration while ${E}^{\Lambda(p)}_{N}(U) =  {E}_{N} (U(\Lambda_1( |\psi_1\rangle \langle \psi_1|)\otimes |\psi_2\rangle \langle \psi_2| )U^\dagger) $ where \(|\psi_1\rangle \otimes |\psi_2\rangle\) is the optimal product state without the noise. We observe that, as is shown in the bottom two figures of Fig. \ref{fig:amp_n_1_u4gen}, the error increases with the increase of noise strength. That is when the noise is weak, the optimal product state in the noiseless case can be used to create almost the same amount of entanglement under the action of the two-qubit unitary operators, although such simplification does not hold in presence of strong noise (see Fig. \ref{fig:amp_n_1_u4gen} for two noise models, ADC and DPC).

\subsection{Modification in hierarchies among unitaries with noise}

As shown in the previous cases, noise can be responsible for diminishing the entangling power of a unitary operator. Let us ask the question of whether the rate of decrease can be different for different classes of unitary operators with variable strength of noise.

To demonstrate it, we Haar uniformly generate unitary matrices in four-dimensional complex Hilbert space. When the input state is influenced by any kind of local noise, we indeed observe that the entangling power can decrease at varying rates with the increase of the noise strength \(p\).   



Let $U_{1}$ and $U_{2}$ be two Haar uniformly generated random unitary matrices.  Suppose their corresponding entangling powers obey the ordering as $\mathcal{E}(U_{1})>\mathcal{E}(U_{2})$. When noise disturbs the inputs, the ordering of the powers can get reversed, i.e., for a finite \(p\), we can have $\mathcal{E}^{\Lambda(p)}(U_{1})<\mathcal{E}^{\Lambda(p)}(U_{2})$ (as shown in Fig. \ref{fig:haar_deph} for the local dephasing noise acted on both the parties of the input state). This is being possible as depending on the noise model and the value of \(p\), the optimal state can also change for a fixed unitary operator. See Appendix \ref{app:mat_hhar}, for the matrix forms of the Haar-uniformly generated unitaries used in Fig. \ref{fig:haar_deph}.



\section{Imperfection in Unitaries: Beyond two-qubits}
\label{sec:imperfection3qubits}

Let us now concentrate on higher-dimensional unitary operators which act on multipartite separable states. The picture becomes more complex when one goes beyond two qubits \cite{Linowski2020, SamirAditi23} since \sm{there are several kinds} of product states in a multipartite domain. For example, in three qubits, \sm{the collection of separable states can be divided into two types} -- fully separable (FS) and biseparable (BS) states \cite{Dur01}. In general, for \(N\)-party systems,  \(N-1\) classes of separable states exist -- starting from the set of fully separable (\(N\)-separable) to the set of \(2\)-separable states \cite{blasone2008}. \sm{In literature, various approaches have been taken to quantify the entangling power of unitary operators in multipartite cases.} Given a unitary operator, the entangling power can be defined as 
 $$\mathcal{E}(U) = \max_{S} \mathcal{E} (U (|\psi_1\rangle \otimes \ldots |\psi_N\rangle )),$$
where \(S\) represents the set of \(N\)-separable states. Notice that in this case, the resulting state may not be genuinely multipartite entangled (GME) (\(1\)-separable) (a pure state is GME when it is not a product in any bipartition). 
As recently noticed by some of us \cite{SamirAditi23}, instead of maximizing over only the set of fully separable states, one can maximize over \(k\)-separable states (\(k=2, \ldots N\)), and the corresponding entangling power of a unitary operator is defined as the maximum genuine multipartite entanglement created from \(k\)-separable states. In other words, the definition is modified as \begin{eqnarray}
    \mathcal{E}(U) = \max_{\mathbb{S}^k} \mathcal{E} (U (|\psi\rangle)),
\end{eqnarray}
\sm{where the maximization is performed over the inputs \(|\psi\rangle\)s  from the set \(\mathbb{S}^k\) where \(\mathbb{S}^k\) is the set of \(k\)-separable states which are not \(k+1\)-separable (for details, see Ref. \cite{SamirAditi23}). Here we note that the sets \(\mathbb{S}^k\) for $k=2,3,\ldots, N$ are disjoint sets.}

While computing the imperfect entangling power of a unitary operator, we also exploit the same definition. We will illustrate that the power behaves similarly as shown in the case of two qubits with respect to the defects present in the system. \sm{However, due to the disjoint sets of optimization, a richer picture emerges. To better visualize the scenario, we will deal with diagonal and Haar uniformly generated unitaries.} 

\subsection{Imperfect entangling power of eight-dimensional diagonal unitaries}
\label{sec:diagonal31ubit}

A diagonal unitary operator in eight-dimensional space can be represented as 
\begin{eqnarray}
    U_{d, 8}^{8}=\text{diag} (e^{i\phi_{1}},e^{i\phi_{2}},e^{i\phi_{3}},e^{i\phi_{4}},e^{i\phi_{5}},e^{i\phi_{6}},e^{i\phi_{7}},e^{i\phi_{8}}),
    \label{eq:diag_8_8}    
\end{eqnarray}
with $\phi_{i} \in (0,2\pi)$  $(i=1,2,\ldots,8)$. Before presenting the results, let us set the stage for analysing multipartite entangling power. To start with, we choose the set of initial states to be either fully separable, given by 
\begin{eqnarray}
    && \ket{\psi^{3-sep}}= \ket{\psi_{1}}\otimes\ket{\psi_{2}}\otimes\ket{\psi_{3}} \nonumber \\
    &&=\otimes_{k=1}^{3}(\cos{\theta_{k}}\ket{0}+\sin{\theta_{k}}e^{i\xi_{k}}\ket{1}),
    \label{eq:state_product_3qubit}
\end{eqnarray}
with $0\leq \theta_{k} \leq \pi$ and $0\leq \xi_{k} \leq 2\pi$ or biseparable state being product in \(12:3\) bipartition, given by
\begin{eqnarray}
     \ket{\psi^{2-sep}}= \ket{\psi_{ij}}\otimes\ket{\psi_{k}}, i\neq j\neq k,\nonumber \\
     = (\sum_{i,j=0}^{1}a_{ij}\ket{ij}) \otimes (\cos{\theta_{k}}\ket{0}+\sin{\theta_{k}}e^{i\xi_{k}}\ket{1}), 
    \label{eq:state_bisep_3qubit}
\end{eqnarray}
where $a_{ij}$ can be parametrized in terms  of $\theta_{1}^{'}$, $\theta_{2}^{'}$, $\theta_{3}^{'}$, and $\xi_{1}^{'}$, $\xi_{2}^{'}$. After the application of \( U_{d, 8}^{8}\), the resulting state is a three-qubit state whose GGM has to be maximized over the set of parameters of the initial state. If the desired value of \(\{\phi_i\}\)s cannot be fixed perfectly, they can be chosen randomly from the Gaussian distribution with mean \(\langle \phi_i \rangle \) and SD, \(\sigma_{\langle \phi_i \rangle}\) \sm{and the quenched average entangling power \(\mathcal{E} (U_{d, 8}^{8} (\langle \phi_i \rangle, \sigma_{\langle \phi_i \rangle}))\) of the given unitary can be computed,  as discussed in Sec.} \ref{subsec:quench}.

\sm{As two extreme cases, \(U_{d,4}^4\) and \(U_{d,1}^4\) have been discussed for two qubits which showed constructive and destructive phenomena  (Sec. \ref{sec:imperfectunitarytwoq}), similar phenomena are  also observed in higher dimension.} Let us first discuss two scenarios where both the benefits and disadvantages of defects are clearly visible.

\begin{figure}
\includegraphics [width=\linewidth]{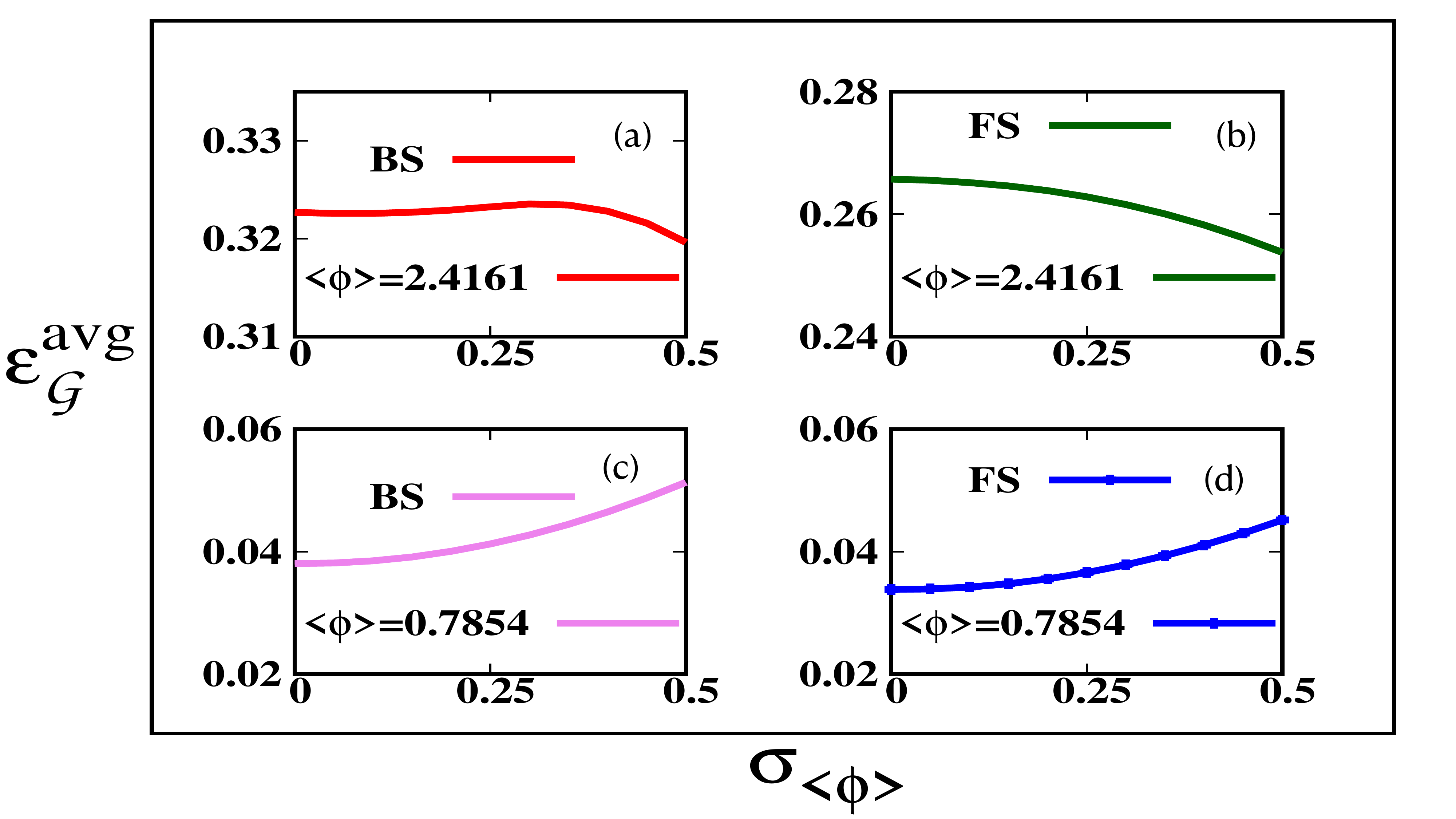}
\caption{\sud{Quenched average entangling power in terms of GGM,  $\mathcal{E}_{\mathcal{G}}^{avg}$ for (\(U^8_{d,1}(\langle \phi \rangle, \sigma_{\langle \phi \rangle})\)) (ordinate) against disorder, $\sigma_{\langle\phi\rangle}$ (abscissa). Here, $\langle \phi \rangle = 2.4161$ (\((a)\) and \((b)\)) and $\langle \phi \rangle = 0.7854$ (\((c)\) and \((d)\)). The maximization involved in the computation of entangling power is done over the set of  biseparable states (BS) in the \(12:3\) bipartition, while FS represents the fully separable input states. The quenched average GGM continues to show nonmonotonic behavior with the increase of disorder in the case of BS and \(\langle \phi \rangle = 2.4161\), as shown in Figs. \ref{fig:nonmonotonic} and \ref{fig:nonmonotonicNeg} in the two-qubit case. The increasing behavior (constructive effect of disorder) is also observed in Fig. \ref{fig:constructive} for two-qubit input states. Both axes are dimensionless.}}
\label{fig:sep_bisep_com}
\end{figure}

{\it Extreme case 1.} When \(\phi_i \equiv \phi\) are all equal (including all vanishing), the entangling power of \(U_{d, 2^N}^{2^N}\) vanishes irrespective of the initial product state. However, when \(\phi_i\)s are chosen randomly with \(\langle \phi_i \rangle = 0\) or {if they are chosen from distinct  Gaussian distributions with the same mean and SD}, we can obtain nonvanishing entangling power of  \(U_{d, 2^N}^{2^N} (\langle \phi \rangle, \sigma_{\langle \phi\rangle})\) which increases with the increase of the disorder strength, \(\sigma_{\langle \phi\rangle}\). Although the exact value of the quenched average GGM  depends on the \(k\)-separable \sm{inputs for different} \(k\), the increasing trend is independent of the kinds of input product states \sud{(see Fig. \ref{fig:sep_bisep_com}(c) and (d) for \(U_{d,1}^{8}(\langle \phi \rangle, \sigma_{\langle \phi \rangle})\)). Note that such an increasing trend in presence of imperfection in the unitary operator is also observed for two-qubits (see Fig. \ref{fig:constructive}).}

{\it Extreme case 2 -- biseparable vs fully separable inputs.} Let us now consider a class of  single-parameter diagonal unitary operators, 
\(U_{d,1}^{2^N}=\text{diag} (1,1, \ldots, e^{i \phi})\). In this case, it is always possible to find a \(2\)-separable optimal state \cite{SamirAditi23}
which leads to the entangling power $\mathcal{E}_{\mathcal{G}}(U_{d,1}^{2^N})=\min[\sin^2{\frac{\phi}{4}},\cos^2{\frac{\phi}{4}}]$. When one chooses with a fixed mean \(\langle\phi\rangle\), \sm{the optimal state that maximizes GGM does not change from the ideal case}. For \(\phi =\pi\), the entangling power achieves its maximum possible value, i.e., \(\mathcal{E}_{\mathcal{G}} (\phi = \pi) =0.5\) for the optimal biseparable state. In contrast to the argument as given in Extreme case 1, \(\mathcal{E}_{\mathcal{G}}^{avg} (\langle\phi =\pi\rangle, \sigma_{\langle \phi\rangle})\) decreases with the increase of \(\sigma_{\langle \phi\rangle}\). 

 Interesting situations arise when the optimization is performed over the set of other \(k\)-separable states, \(k\neq 2\). For example, in the three-qubit disorder-free case, taking the optimal  fully separable input  state, 
\(     \ket{\psi^{3-sep}_{opt}}= (0.60922\ket{0}+0.793001\ket{1})^{\otimes 3}\),
      we obtain $\mathcal{E}_{\mathcal{G}}(U_{d,1}^{8}(\phi = \pi)) =0.34$.  We observe that  $\mathcal{E}_{\mathcal{G}}^{avg} (U_{d,1}^{8}(\langle \phi \rangle = \pi, \sigma_{\langle \phi\rangle}))$ decreases with the increase of \(\sigma_{\langle \phi\rangle}\) although the rate of fall in the case of the fully separable state is much slower than that of the biseparable states.

{\it Nonmonotonic case 3. } As also illustrated in the case of two-qubit entangling unitaries, let us concentrate on the diagonal unitaries which may manifest nonmonotonic behavior with respect to the increase of disorder. We choose a single-parameter family of unitary, \(U_{d,1}^8 (\langle \phi\rangle)\) with the same value of \(\langle \phi\rangle\) as chosen in Fig. \ref{fig:nonmonotonic}. Let us now illustrate the behavior of this diagonal unitary operator when chosen from Gaussian distribution with mean \(\langle \phi \rangle = 2.4161\) and varying SD (see Fig. \ref{fig:sep_bisep_com}). \sud{We observe that when the optimal input state is biseparable in the \(12 : 3\) bipartition, \(\mathcal{E}^{avg}_{\mathcal{G}}\) behaves nonmonotonically with \(\sigma_{\langle \phi\rangle}\) while monotonic (non-increasing) character is observed when the optimal input state is fully separable}. 

\sud{\subsubsection{Imperfections in unitaries for three qubits and beyond}}

\begin{figure}
\includegraphics [width=\linewidth]{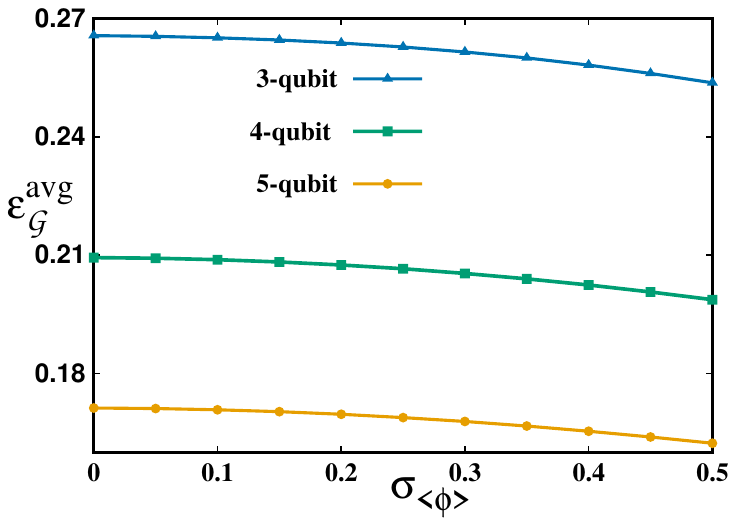}
\caption{\sud{Quenched average entangling power in terms of GGM,  $\mathcal{E}_{\mathcal{G}}^{avg}$, (ordinate) for diagonal unitary operators consisting of a single parameter in eight, sixteen and thirty-two-- dimensional space against disorder strength, $\sigma_{\langle\phi\rangle}$ (abscissa) with a fixed mean, $\langle \phi \rangle = 2.4161$. Here, the maximization to obtain the entangling power is performed over the set of fully separable three-, four- and five-qubit input states. The quenched average GGM continues to show the same decreasing behavior with the increase of dimension. Both axes are dimensionless.}}
\label{fig:decrease_4_5}
\end{figure}

\sud{Extending our analysis to larger systems, we perform numerical simulations for computing entangling powers of a single-parameter diagonal unitary operators in \(2^4\)- and \(2^5\)- dimensional space where the optimization is done over the set of fully separable states. Our results indicate that the behavior observed in the three-qubit case  also persists in these higher-dimensional scenarios (see Fig. \ref{fig:decrease_4_5}). Specifically, both \(\mathcal{E}_{\mathcal{G}}^{avg}(U_{d,1}^{16}(\langle \phi \rangle,\sigma_{\langle \phi \rangle}))\) and \(\mathcal{E}_{\mathcal{G}}^{avg}(U_{d,1}^{32}(\langle \phi \rangle,\sigma_{\langle \phi \rangle}))\) decrease with increasing \(\sigma_{\langle \phi\rangle}\), when the optimization is performed over the set of four- and five- qubit fully  separable input states, respectively. Moreover, these observations possibly suggest that we can also obtain increasing and nonmonotonic behavior in entangling powers for higher-dimensional unitary operators if the sets for optimizations and \(\langle \phi \rangle\) are chosen suitably as found in Fig. \ref{fig:sep_bisep_com}.  }

\subsection{Ranking among noisy entangling power}

Let us now study the entanglement of the unitary operators where all the qubits are affected by three paradigmatic local noise models. First of all, \sm{such a study requires quantification of entanglement in mixed states between three or more parties \cite{Horodecki2009} about which only limited knowledge is available. Here we adopt the concept of monogamy of entanglement to assess the entangling power of such unitary operators }\cite{ckw00,osborne06,ou07,monorev17}. In particular, we compute the monogamy score of squared negativity for an arbitrary \(N\)-party resulting state, \(\rho_{1\ldots N}\), given by  \(\delta_{N^2} = N^2(\rho_{2:13\ldots N}) - \sum_{i=1, i\neq 2}^N N^2(\rho_{2i})\) where \(\rho_{2i}\) represents the reduced density matrices of \(\rho_{12\ldots N}\) and \(N^2(\rho_{2:13\ldots N})\) represents the \sm{square of the negativity of \(\rho_{1\ldots N}\) in the bipartition \( 2 :\text{rest}\) \sh{with \(2\) being the nodal observer}. 
We now examine the effects of noise on the entangling powers of the diagonal, permutation, and Haar uniformly generated eight-dimensional unitary operators. }

{\it Diagonal unitaries.} In case of \(U_{d,1}^8(\phi)\), it is observed that the optimal biseparable state in the \(12:3\) bipartition, $\ket{\psi^{2-sep}_{opt}}= \frac{1}{\sqrt{2}}(\ket{00}+\ket{11})\frac{1}{\sqrt{2}}(\ket{0}+\ket{1})$ remains same for the PDC and DPC channels for the entire range of noise strength while for the ADC channel, the optimal inputs may vary with \(p\).  
 
\sm{However},  when maximization is performed over the set of fully separable inputs, the optimal state always changes with the strength of noise for all the three noisy channels. 

\begin{figure}
\includegraphics [width=\linewidth]{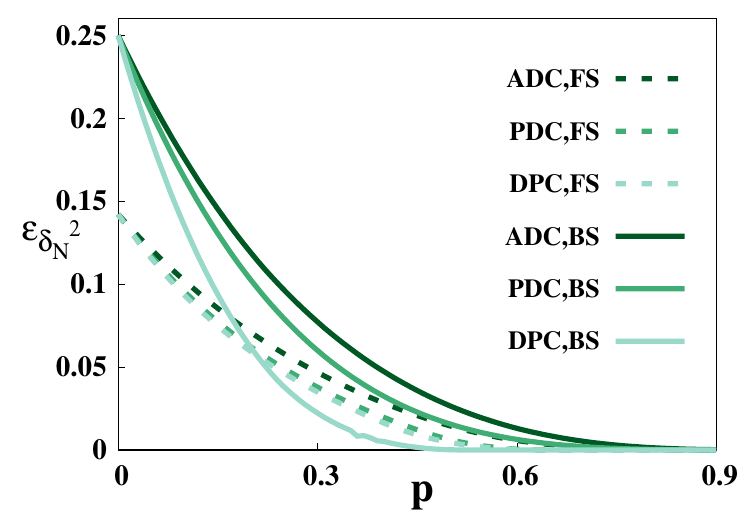}
\caption{Entangling power via squared negativity monogamy score, $\mathcal{E}_{{\delta}_{N^{2}}}$ (y-axis)  against local noise parameter, $p$ (x-axis) corresponding to diagonal unitary, $U_{d,1}^8(\phi)$ with $\phi = \pi$. The dashed line corresponds to the set of FS (fully separable) states while the solid line is for the set of  BS (biseparable) states. Qubits are affected by ADC, PDC, and DPC (sequentially colored from dark green to light green). Interestingly, notice that in the noiseless case, the entangling power obtained from the biseparable optimal input state is higher than that of the FS ones while for moderate \(p\) in the DPC, the optimal state changes from biseparable to fully separable ones.  Both the axes are dimensionless.}
\label{fig:u8_1_noise}
\end{figure}

Like four-dimensional \sm{diagonal unitary, we also observe that the fastest rate in which the entangling power vanishes for both separable and biseparable inputs, occurs in the case of depolarizing channel} while \(\mathcal{E}_{\delta_{N^2}}\) decreases slowly when local ADC disturbs the inputs. Again, numerical investigation suggest that the entangling power of the  diagonal unitary operator follows the hierarchy as
\begin{eqnarray} 
\mathcal{E}_{\delta_{N^2}}^{\text{DPC}}>
\mathcal{E}_{\delta_{N^2}}^{\text{PDC}} > \mathcal{E}_{\delta_{N^2}}^{\text{ADC}}, \forall p>0
\end{eqnarray}
which is true for both fully separable and biseparable inputs. \sud{Instead of considering biseparable inputs which is product in the \(12:3\) bipartition, if one takes product in other bipartition, similar results may emerge although in \(\delta_{N^2}\), the nodal point may have to be changed.} Again, \sm{the local ADC turns out to be the best in terms of the robustness of entangling power of diagonal unitary operators}. Moreover, for a single-parameter diagonal unitary operator, typically biseparable states are more favorable inputs than the fully separable states with respect to the creation of genuine multipartite entanglement in the noiseless case. We find that when amplitude and phase damping channels with a fixed \(p\) act on the inputs,  biseparable inputs remain better than the fully separable ones although it is no more true for the depolarizing channel after a certain \(p\) value. For example, \(U_{d,1}^8(\phi = \pi)\), we find a threshold noise strength, \(p\), above which \sm{fully separable} states can produce a higher amount of entanglement compared to the case of biseparable 
inputs (see Fig. \ref{fig:u8_1_noise}). 

\begin{figure}
\includegraphics [width=\linewidth]{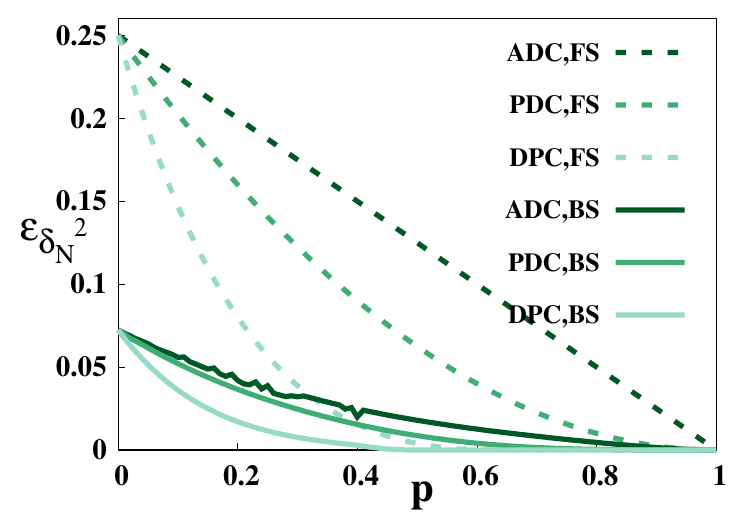}
\caption{Entangling power via squared negativity monogamy score, $\mathcal{E}_{{\delta}_{N^{2}}}$  (vertical axis)  against local noise parameter, $p$ (horizontal axis) corresponding to permutation unitary, $\mathbb{P}(1,7)$.  All other specifications, including color coding and
line types, are the same as in Fig. \ref{fig:u8_1_noise}. Both the axes are dimensionless.}
\label{fig:permutation_noise}
\end{figure}

{\it Permutation unitary operators. } We now choose an example of a permutation unitary operator to highlight that biseprable inputs do not always lead to higher entangling power than the \sm{fully separable} ones both in the absence and presence of noise. For example, \(\mathbb{P}(1,7)\) can continue to create a higher multipartite entangled output state from \sm{fully separable} states compared to the case when the optimization is performed over the set of biseparable states \sh{in the \(12:3\) bipartition}, as depicted in Fig. \ref{fig:permutation_noise}. However,   there are permutation unitary operators (as listed in Ref. \cite{SamirAditi23}) which can create higher GME states from biseparable inputs than the \sm{fully separable} ones \cite{SamirAditi23}.  Further, we notice that the response of noisy channels found in the case of diagonal \sm{unitary operators} remains the same for the transposition unitary operators also.  Specifically, the entangling power of the transposition \sm{unitary operators} manifests maximum robustness in the presence of ADC for all kinds of input separable states (see Fig. \ref{fig:permutation_noise}).

\begin{figure}
\includegraphics [width=\linewidth]{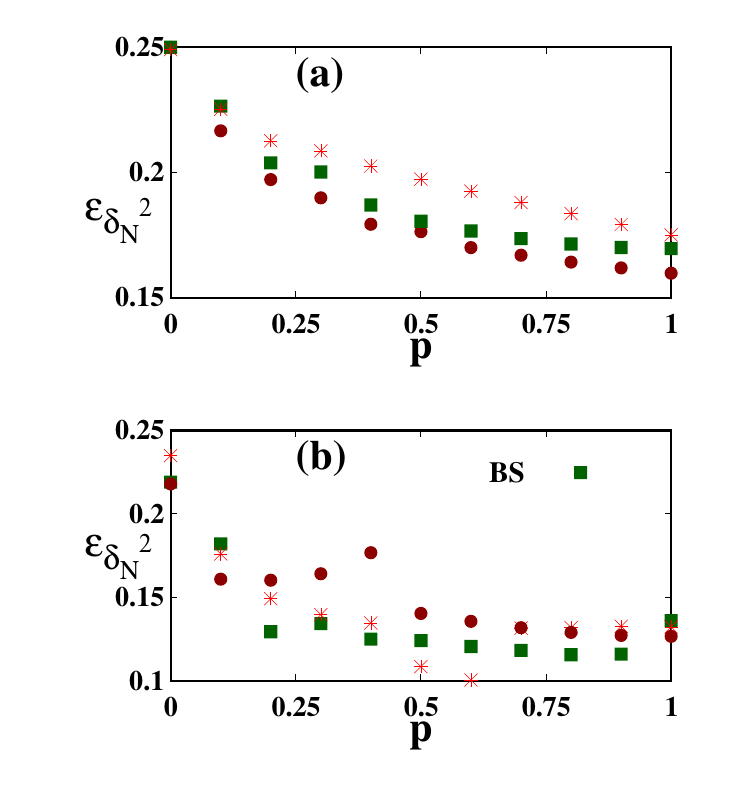}
\caption{Entangling power in terms of monogamy score, $\mathcal{E}_{{\delta}_{N^{2}}}$ (vertical axis) against noise parameter, $p$ of PDC channel (horizontal axis)  corresponding to Haar uniformly generated eight-dimensional unitaries. (a) The optimization is performed over the set of FS states while in (b), the same optimization is performed over the set of biseparable states. As depicted in Fig. \ref{fig:haar_deph},  it again illustrates that the ordering among unitaries changes in higher-dimensional systems also.   Both the axes are dimensionless.}
\label{fig:haar_3qubit}
\end{figure}

{\it Haar uniformly generated unitary operators.} Let us now  Haar uniformly  generate eight-dimensional unitary operators which act on the set of fully separable and biseparable states. We compute the entangling power in terms of the monogamy score of squared negativity for the resulting states. As depicted in Fig. \ref{fig:haar_deph}, \sm{in the case of four-dimension}, we again highlight that for a given noise strength \(p\) and the fixed noise model, the decrease \sm{in the entangling power} for two fixed unitary operators can happen at a different rate (see Fig. \ref{fig:haar_3qubit}). It implies that when the noise present in the circuit is known, it is quite \sm{plausible} that we have to design the quantum circuit composed of different gates that can produce the desired quantum resource required for a certain task in a \sm{different way from the one that is suitable} in the noiseless case. 

\section{Conclusion}
\label{sec:conclu}

The concept of entanglement of operators parallels the idea of entanglement content in states. Here, the entangling power of a fixed unitary operator is measured by the maximum entanglement generated when it acts on an optimal initially unentangled state. Essentially,  the optimal product state is the one that, under the influence of the given unitary operator, yields the highest amount of entanglement among all possible product states.

Ideally, since the isolated systems cannot be prepared, it is necessary to assume that the system is always in contact with the environment. Similarly,  the implementation of unitary operators cannot be perfect. We incorporated imperfections and noise in the framework of entangling power which we call as imperfect (noisy) entangling power. Specifically, when the device implementing a unitary operator cannot be calibrated to a desired value, it takes the values randomly from a Gaussian distribution with an intended value as mean, and depending on the device's quality, standard deviation may vary. In this situation, instead of entangling power, one has to compute quenched average entangling power with the averaging being performed over several realizations of the chosen values. \sud{Our numerical simulations up to five qubits  indicate that the quenched average entangling power decreases with the increase of disorder strength, although in some nontrivial situations, it can increase as well  behaves nonmonotonically. Further, for generic two-qubit unitaries, we found that the quenched average entangling power becomes independent of the parameters, which is not the case for the ideal scenario. Both the results highlight the positive impacts of disorder on entangling power, as  reported in the literature of many-body physics \cite{PhysRevLett.62.2503,disorderBerarakshit14,Pritam_2022}.}

On the other hand, when noise disturbs the input states, we first demonstrated that the optimal input states in the noiseless case which generate maximal entanglement upon the action of unitary operators change with the noise model and the strength of noise. In all the scenarios and noise models considered in this work, we exhibited that entangling power displays higher resilience against amplitude damping (AD) noise compared to phase damping (PD) and depolarizing (DP) noise.   Furthermore, we found that the ranking of unitary operators based on their entanglement capabilities in noiseless conditions can alter significantly in the presence of noise.  It is known that in the noiseless case, some classes of unitary operators can create higher multipartite entanglement from the biseparable optimal input states compared to the fully separable optimal ones. We found that in the presence of noise, those unitary operators acting on the optimal fully separable input states can now produce higher multipartite entanglement than that obtained from the optimal biseparable inputs.

These results underscore the intricate interplay between imperfections in unitaries, noise present in the input states, and their collective impact on entanglement generation via quantum gates. Understanding these complexities is vital for optimizing quantum protocols and designing robust quantum systems in laboratories.



\acknowledgements
 The authors acknowledge the support from the Interdisciplinary Cyber Physical Systems (ICPS) program of the Department of Science and Technology (DST), India, Grant No.: DST/ICPS/QuST/Theme- 1/2019/23 and the use of cluster computing facility at the Harish-Chandra Research Institute.  

 \appendix

 \section{Prerequisites: Tools used to quantify Entanglement}
 \textit{Multiparty entanglement measure: Generalized Geometric Measure.}
Genuine multiparty entangled pure states are those which are not products of any bipartition. Several multipartite entanglement measures have been proposed and proved to be crucial for quantum information protocols, although they are not always easy to compute \cite{Horodecki2009}. The generalized geometric measure \cite{aditi2010}, $\mathcal{G}$, is a distance-based measure, defined as the distance of a given state from the set of non-genuine multiparty entangled states. In this paper, we choose a GGM to quantify GME content in the resulting multipartite states obtained after the action of the unitary operator on product states. It is defined for an arbitrary multipartite state, \(|\psi\rangle\) as 
\begin{eqnarray}
   && \mathcal{G}(\ket{\psi})= 1 - \max_{\phi \in nG} |\langle \phi | \psi\rangle |^2 = 1 - \max [\{e_{i}\}]
\end{eqnarray}
where $nG$ denotes the set of all nongenuinely
multipartite entangled states, \(|\phi\rangle\) and $\sqrt{e_{i}}$ is  the Schmidt coefficients of all the bipartitions of \(|\psi\rangle\).

\textit{Negativity of a bipartite state.}
During the presence of imperfection, pure states become mixed and we will use negativity to measure the entanglement of the output state. Negativity based on the partial transposition criteria \cite{Peres96, Horodecki96} on the composite Hilbert space of $\mathcal{H}_1^{d}\otimes\mathcal{H}_2^{d}$ is defined  for an arbitrary bipartite state, \(\rho_{12}\), as \cite{Vidal02}  
\begin{eqnarray}
    N_{12}\equiv N(\rho_{12})&=&\frac{||\rho_{12}^{T}||_{1}-1}{2},
    \label{eq:negativity}
\end{eqnarray}
where $||\rho||_{1}=\text{tr}(\rho\dag \rho)$ is the trace norm. It reduces to 
\begin{eqnarray}
    N(\rho_{12})&=&\sum_{j}|\lambda_{j}^{n}|,
    \label{eq:negativity_1}
\end{eqnarray}
where $\lambda_{j}^{n}$'s represent the negative eigenvalue of $\rho_{12}^{T}$ with superscript ``$n$" denoting the negative eigenvalues of the partial transposed state.

\textit{Monogamy Score.} Since we will be dealing with noisy multisite entangled states, computable entanglement measures are limited in the literature. However, if we are interested in how quantum correlations (QC) can be distributed (shared) among many nodes, a natural choice is to adopt the concept of monogamy of entanglement \cite{ckw00}. It states that in a multipartite state, if two parties are maximally quantum correlated, other parties cannot share any quantum correlation with these two. Beyond this extreme scenario, it provides an upper bound on the sum of entanglements that a party can share with other sites. Mathematically, for a given multiparty QC measure $N$, and a n-party state $\rho_{1 2 \ldots k \ldots n}$ with $k$ being the nodal observer, the monogamy score for $N$ reads as 
\begin{eqnarray}
    \delta_{N^2}(\rho_{1 \ldots n})= N^2(\rho_{k:\text{rest}})-\sum_{i=1,i\neq k}^n N^2(\rho_{ki}).
    \label{eq:monogamy}
\end{eqnarray}
where $\rho_{k:\text{rest}}$ and $\rho_{ki}$ represent the multiparty state in the bipartition, $k:\text{rest}$ with rest being  all the parties except party $k$ and the bipartite reduced density matrix between the party $k$ and $i$ respectively. In our work, we choose squared the negativity $N^2$ which is defined as the sum of the absolute negative eigenvalues of the partially transposed state.



 \section{Calculation of  entangling power of diagonal unitary operator in terms of negativity}
 
Let us consider a diagonal unitary operator, \(U_{d,1}^4(\phi)\). When it acts on the arbitrary two-qubit product state, the entangling power takes the form as 
\begin{eqnarray}
   \mathcal{E}_{N}(U_{d,1}^{4}) &=&\max_{\theta_{1},\theta_{2},\xi_{1},\xi_{2}}  \mathcal{E}_{N}\left(U_{d,1}^{4} (\otimes_{k=1}^{2}\cos\theta_{k}\ket{0}+\sin\theta_{k}e^{i\xi_{k}} \ket{1})\right).\nonumber\\
    \label{eq:state_u4_1_negativity}
\end{eqnarray}
The resulting density matrix required to calculate negativity, $N$, is given by 
\begin{widetext}
\begin{eqnarray}
\rho&=&
\left(
\begin{array}{cccc}
  \cos ^2 \theta _k \cos ^2 \theta _2 & A & B & C \\
 A^* & \sin ^2 \theta _2 \cos ^2 \theta _1 & D & E \\
 B^* & D^* & \sin ^2 \theta _1 \cos ^2 \theta _2  & F \\
 C^* & E^* & F^* & \sin ^2 \theta _1 \sin ^2 \theta _2 \\
\end{array}
\right),
\end{eqnarray}
\end{widetext}
where $A=e^{-i \xi _2} \sin \theta _2 \cos ^2 \theta _1 \cos \theta _2$, \\
$B=e^{-i \xi _1} \sin \theta _1 \cos \theta _1 \cos ^2 \theta _2$, \\
$C=e^{-i \xi _1-i \xi _2-i \phi } \sin \theta _1 \sin \theta _2 \cos \theta _1 \cos \theta _2 $,\\
$D=e^{-i \xi _1+i \xi _2} \sin \theta _1 \sin \theta _2 \cos \theta _1 \cos \theta _2$, \\ 
$E=e^{-i \xi _1-i \phi } \sin \theta _1 \sin ^2 \theta _2 \cos \theta _1 $,\\
and
$F=e^{-i \xi _2-i \phi } \sin ^2 \theta _1 \sin \theta _2 \cos \theta _2 $.
After optimizing \(\theta_k\)s, we find that the negativity reads as
 \begin{eqnarray}
 N&=&\frac{1}{16} \big(4 | -1+e^{i \phi }| +| 2 \sqrt{e^{i \phi } (1+e^{i \phi })^2}+4 e^{i \phi }| \nonumber\\&+&| 4 e^{i \phi }-2 \sqrt{e^{i \phi } (1+e^{i \phi})^2}| -8\big). 
 \label{eq:negativity_u_4_1}
 \end{eqnarray}
Like GGM, the entangling power, in this case also gets maximized when \(\theta_k = \pi/4\) \(\forall \xi_k\)s (\(k=1,2\)). 

\section{The matrix forms corresponding to Haar random generated unitaries.}
\label{app:mat_hhar}

The explicit matrix forms of Haar randomly generated unitaries used in Fig. \ref{fig:haar_deph} are given by
\begin{widetext}
\begin{eqnarray}
U_1&=&
\left(
\begin{array}{cccc}
-0.1328 + 0.2848i & +0.3609 - 0.0663i & +0.0975 - 0.3560i & -0.3627 - 0.7063i \\
+0.1582 + 0.4851i & -0.2756 - 0.2412i & +0.2530 - 0.6070i & +0.3639 + 0.2016i \\
-0.3395 + 0.1700i & -0.6838 - 0.3707i & -0.1359 + 0.3228i & -0.0112 - 0.3579i \\
-0.7048 + 0.0060i & -0.0797 + 0.3463i & -0.3833 - 0.4022i & -0.0754 + 0.2502i \\
\end{array}
\right),
\end{eqnarray}
\begin{eqnarray}
U_2&=&
\left(
\begin{array}{cccc}
-0.3105 - 0.4472i & -0.2782 + 0.2450i & +0.2835 + 0.3510i & -0.6021 - 0.0030i \\
+0.4682 - 0.0869i & -0.5794 + 0.1892i & -0.1666 - 0.0687i & +0.0463 + 0.6059i \\
-0.0522 - 0.6011i & +0.2999 - 0.6057i & -0.0134 - 0.0419i & +0.0555 + 0.4173i \\
-0.3121 - 0.1243i & +0.0490 + 0.1786i & -0.7670 - 0.4167i & -0.3008 - 0.0207i \\
\end{array}
\right),
\end{eqnarray}
\begin{eqnarray}
U_3&=&
\left(
\begin{array}{cccc}
+0.1515 - 0.0973i & +0.1520 + 0.2570i & +0.1770 - 0.7953i & -0.0386 - 0.4617i \\
+0.0142 + 0.4815i & +0.0317 + 0.7051i & -0.4129 + 0.0758i & +0.3059 - 0.0082i \\
+0.5406 + 0.5763i & -0.3954 - 0.2356i & -0.0320 - 0.0407i & -0.3844 - 0.1154i \\
-0.3255 - 0.0724i & -0.1310 + 0.4286i & +0.2146 + 0.3336i & -0.6472 - 0.3344i \\
\end{array}
\right),
\end{eqnarray}

\begin{eqnarray}
U_4&=&
\left(
\begin{array}{cccc}
+0.4815 + 0.2728i & -0.1891 + 0.2117i & +0.6345 - 0.4073i & +0.0803 - 0.1952i \\
+0.3253 - 0.2366i & -0.6629 - 0.2381i & -0.1767 + 0.2749i & -0.2700 - 0.4031i \\
-0.2068 - 0.2555i & +0.4632 + 0.0266i & +0.3072 + 0.1931i & -0.0981 - 0.7317i \\
-0.5813 + 0.2931i & -0.3827 + 0.2482i & -0.2519 - 0.3599i & +0.1469 - 0.3918i \\
\end{array}
\right),
\end{eqnarray}
\begin{eqnarray}
U_5&=&
\left(
\begin{array}{cccc}
+0.0126 - 0.3590i & +0.0075 + 0.5262i & +0.1540 - 0.6068i & -0.3297 + 0.3056i \\
+0.8058 + 0.3372i & +0.0757 + 0.2171i & -0.3882 - 0.1587i & +0.0588 - 0.0688i \\
-0.0408 - 0.2034i & +0.0895 + 0.2377i & +0.2034 - 0.2737i & +0.6738 - 0.5676i \\
+0.0709 - 0.2447i & -0.5141 - 0.5843i & -0.2809 - 0.4869i & +0.1134 + 0.0255i \\
\end{array}
\right).
\end{eqnarray}
\end{widetext}


\bibliography{ref}

\end{document}